\documentclass[aps,nofootinbib,preprint]{revtex4-1}

\usepackage{verbatim}
\usepackage[T1]{fontenc}
\usepackage[utf8]{inputenc}
\usepackage[american]{babel}
\usepackage{epsfig}
\usepackage{graphicx,subcaption,caption}
\usepackage{booktabs}
\usepackage{multirow}
\usepackage{dcolumn}
\usepackage{amsmath}
\usepackage{mathtools}
\usepackage{amsfonts}
\usepackage{amssymb}
\usepackage{ulem}
\usepackage{epstopdf}
\usepackage{bm}
\usepackage{braket}
\usepackage{enumitem}
\usepackage{soul}
\usepackage[table]{xcolor}
\usepackage{color}
\usepackage{transparent}
\usepackage{pifont}
\usepackage{enumitem}
\usepackage{rotfloat}

\definecolor{navyblue}{rgb}{0.0, 0.0, 0.5}
\definecolor{royalblue}{rgb}{0.25, 0.41, 0.88}
\definecolor{cadmiumgreen}{rgb}{0.0, 0.42, 0.24}
\definecolor{blue-violet}{rgb}{0.54, 0.17, 0.89}
\definecolor{darkviolet}{rgb}{0.58, 0.0, 0.83}
\definecolor{orange(colorwheel)}{rgb}{1.0, 0.5, 0.0}

\usepackage{hyperref}
\hypersetup{
    colorlinks=true, 
    linkcolor=royalblue, 
    citecolor=magenta}

\usepackage{booktabs}
\usepackage{multirow}
\usepackage{dcolumn}
\usepackage{colortbl}

\begin{document}

\title{Redshift-space distortion constraints on the neutrino mass and models to alleviate the Hubble tension}

\author{Yo Toda}
\email{y-toda@particle.sci.hokudai.ac.jp}
\affiliation{Department of Physics, Hokkaido University, 
Sapporo 060-0810, Japan \looseness=-1}

\author{Osamu Seto}
\email{seto@particle.sci.hokudai.ac.jp}
\affiliation{Department of Physics, Hokkaido University, 
Sapporo 060-0810, Japan \looseness=-1}

\begin{abstract}
We discuss the neutrino mass and Hubble tension solutions and examine their effects on the redshift-space distortion (RSD) observations. An analysis with RSD data indicates smaller amplitude of perturbation. Including RSD data results in a slightly weaker upper limit on the neutrino mass than that derived for data without RSD, which is common in other extended models too. We have evaluated the impacts of RSD observations on some extended models, including the varying electron mass model, a time-dependent dark energy model with two parameter equations of state (EOS), and a model where the number of neutrino species is free. When we estimate the cosmological parameters for data including RSD, we found that the EOS parameter for dark energy is larger than that of the cosmological constant, and the effective number of neutrino species is smaller than the standard value, which infers a smaller present Hubble parameter $H_0$. From the viewpoint of cosmological tensions,  the varying electron mass model with nonzero neutrino mass option looks promising to relax the Hubble tension and the $S_8$ tension simultaneously.
\end{abstract}
\preprint{EPHOU-24-016}

\maketitle

\section{Introduction}
\label{sec:introduction}

The standard cosmology with cold dark matter (CDM) and cosmological constant $\Lambda$, called the $\Lambda$CDM model, is a great way to explain the large scale properties of our Universe. The matter power spectrum, also known as the galaxy two-point correlation function, is the primary quantity that describes the properties of large scale structures. This can provide us with information about the growth of the density perturbation. The redshift of galaxies is measured by galaxy surveys. The positions of galaxies differ at the redshift-space and real position actually, which is referred to as redshift space distortions (RSD)~\cite{Kaiser:1987qv,Hamilton:1992zz}. This distortion originates from not only the cosmic Hubble expansion but also the peculiar velocity of an individual galaxy.

By measuring the RSD, we may find the growth rate of the matter fluctuation~\cite{Song:2008qt,Blake:2011rj,Blake:2013nif,Simpson:2015yfa,Okumura:2015lvp,Gil-Marin:2016wya,Mohammad:2018mdy,Said:2020epb,eBOSS:2020gbb,Avila:2021dqv}, as a function of redshift $z$. To be precise, it is expressed by the value $f(z)\sigma_8(z)$, where 
 $f\equiv d\ln \delta_m/d\ln a$ is the growth rate, $\delta_m(z)\equiv \delta \rho_m(z)/\bar{\rho}_m(z)$ is the density contrast, $a$ is the scale factor of our Universe, $\sigma_8$ is the amplitude of mass fluctuations in spheres of radius  $8h^{-1}~\mathrm{Mpc}$, and $h$ is the dimensionless Hubble parameter of $H_0 = 100 h $ km$/s/$Mpc. The growth rate reported in those observations is lower than that inferred by Planck~\cite{Planck:2018vyg} in the CDM model. Numerous extended models have been proposed to explain the discrepancy. See, for instance, Refs.~\cite{Joudaki:2017zdt,Gomez-Valent:2017idt,Nesseris:2017vor,Gomez-Valent:2018nib,Benisty:2020kdt,Wright:2020ppw, Nunes:2021ipq,Briffa:2023ozo,Nguyen:2023fip,Adil:2023jtu,Tang:2024gtq,Sabogal:2024yha,Toda:2024fgv} for the recent studies.

Another cosmological tension on the standard $\Lambda$CDM model under debate is so-called the Hubble tension, which is discrepancy of the present Hubble parameter $H_0$ measured by local observations~\cite{Riess:2021jrx,Riess:2023bfx,Riess:2024ohe,Wong:2019kwg,Freedman:2019jwv,Freedman:2020dne,Freedman:2021ahq} and inferred from cosmological observations such as cosmic microwave background radiation (CMB)~\cite{Planck:2018vyg} and baryon acoustic oscillation (BAO)~\cite{Beutler:2011hx,Ross:2014qpa,BOSS:2016wmc}. 
While no satisfactory cosmological model has been proposed yet, some models can alleviate the Hubble tension significantly. For a recent review, see e.g., Refs.~\cite{DiValentino:2021izs,Perivolaropoulos:2021jda,Schoneberg:2021qvd,Shah:2021onj,Abdalla:2022yfr,Hu:2023jqc}.

In this paper, we examine simple extensions of $\Lambda$CDM and models that have been proposed to alleviate the Hubble tension and evaluate their effects and constraints from the RSD data.
This paper is organized as follows. After we describe the models to be discussed in this paper in the Sec.~\ref{sec:models}, we explain the method and datasets of our analysis in the Sec.~\ref{sec:data} and present the results in the Sec.~\ref{sec:results}.
We summarize this paper in the Sec.~\ref{sec:conclusions}.

\section{Models}
\label{sec:models}

In this section, we give an explanation of the four models we examine and discuss their effects on the observations. We consider two different options for neutrino mass when considering each model. One is to fix the neutrino mass, while the other is to vary the sum of neutrino masses $\sum m_{\nu}$. 

\subsection*{Neutrino mass}
\label{subsec:mnu}
When we fix the neutrino mass, we assume the smallest neutrino mass in the spectrum of normal mass ordering $m_{\nu}= (0, 0, 0.06)~\mathrm{eV}$. On the other hand, when we vary the neutrino mass, we vary the total neutrino mass $\sum m_{\nu}$ with keeping the normal mass ordering and the neutrino energy density given as $\Omega_{\nu}h^2  = \sum m_{\nu}/(94.07\, \mathrm{eV}) \times(3.046/3)^{3/4}$ in the standard ($N_{\mathrm{eff}}=3.046$) case.
As the neutrino mass increases, the growth rate of the matter fluctuations with a wave number larger than the free-streaming wave number is suppressed (See Refs.~\cite{KATRIN:2019yun, Oldengott:2019lke, Pan:2015bgi, Tanseri:2022zfe, Boyle:2017lzt, Allali:2024aiv,Loverde:2024nfi} for the recent works of the neutrino mass and large scale structure and 
Refs.~\cite{TopicalConvenersKNAbazajianJECarlstromATLee:2013bxd, Lesgourgues:2006nd} for the review).
Since the lower growth rate $f\sigma_8$ are reported than assuming $\Lambda$CDM best-fit values, we expect that a larger neutrino mass is preferred from the RSD data 
\footnote{When we focus the RSD data only from SDSS-IV~\cite{eBOSS:2020yzd} which has reported a larger $f\sigma_8$, a lower neutrino mass seems to be preferred~\cite{DiValentino:2021hoh}. There are also
recent apparent unphysical indication of $\sum m_{\nu}<0$ from the DESI BAO~\cite{DESI:2024mwx,Craig:2024tky}.}.

\subsection{$\Lambda$CDM}
\label{subsec:lcdm} 

The standard cosmological model has the six free parameters:
the baryon density $\omega_b=\Omega_b h^2$, the CDM density $\omega_c=\Omega_c h^2$, the amplitude of the primordial density perturbation $A_s$, the spectral index of the primordial density perturbation $n_s$, the acoustic scale  $\theta_* = r_* / D_M$ with $r_*$ the sound horizon at recombination and $D_M$  the angular diameter distance to the last scattering surface, and the optical depth of the reionization $\tau$.

\subsection{Varying electron mass}
\label{subsec:me} 

The varying electron mass model~\cite{Barrow:2005qf} has one additional parameter: $m_e/m_{e0}$, where $m_e$ is the electron mass in the recombination era and $m_{e0}=0.511~\mathrm{MeV}$ is the current electron mass. In this paper, we consider a scenario where $m_e/m_{e0} > 1$ at the early Universe and the value of electron mass changes to the current value after the recombination is complete.

\begin{figure}
\includegraphics[width=16cm]{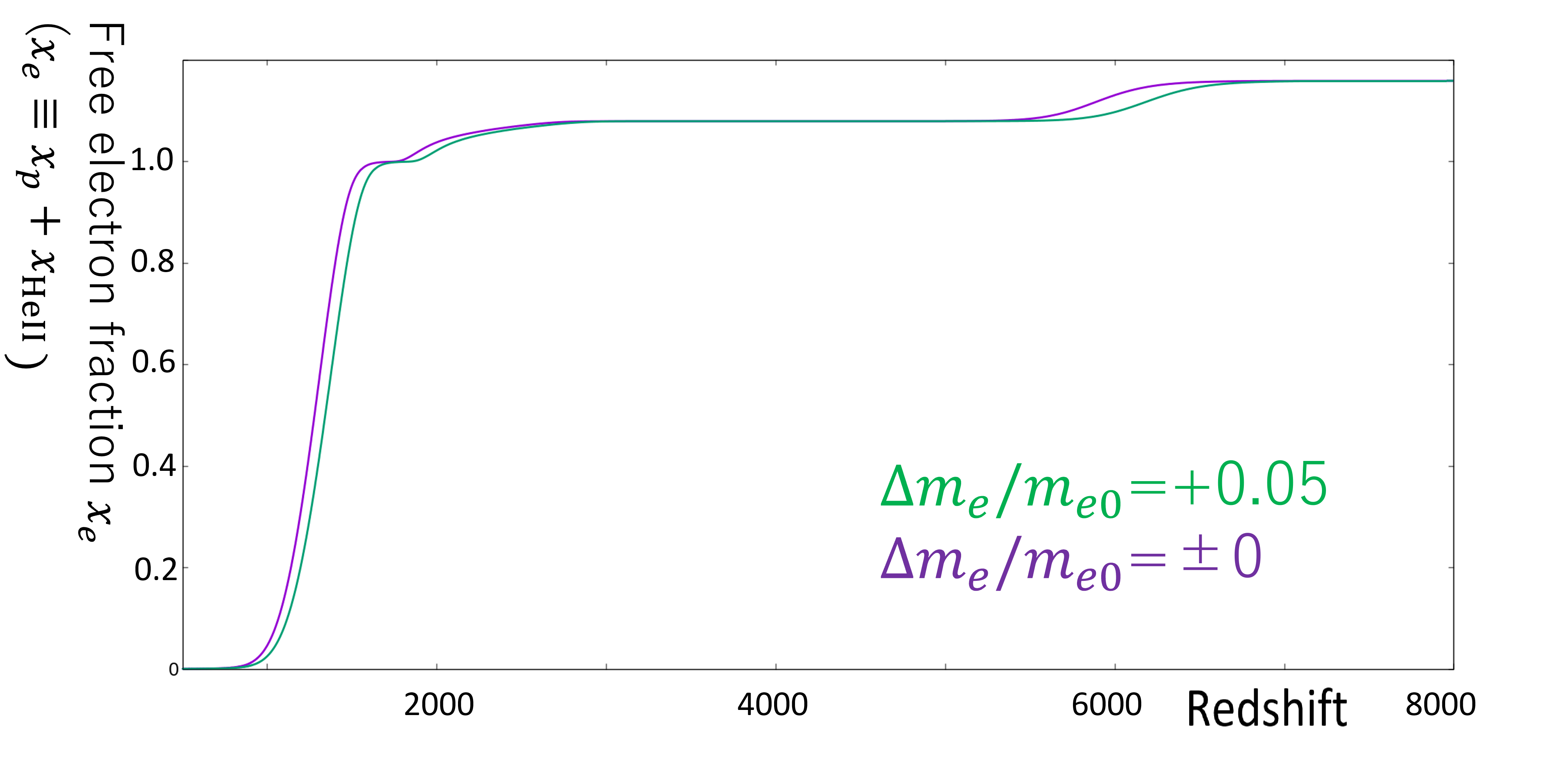}
\caption{Recombination.}
\label{fig:rec} 
\end{figure}

The varying electron mass model is a model that is highly promising for relaxing the Hubble tension~\cite{Hart:2017ndk,Hart:2019dxi,Sekiguchi:2020teg,Schoneberg:2021qvd} and has been studied from various aspects~\cite{Seto:2022xgx,Hoshiya:2022ady,Solomon:2022qqf,Seto:2024cgo,Baryakhtar:2024rky}. As is explained in the Planck(2015)~\cite{Planck:2014ylh}, the main contributions of a larger electron mass are the higher energy levels of hydrogen and Lyman alpha photons,
both of which are directly proportional to the electron mass\footnote{In fact, we also include the minor corrections: Thomson scattering rate, the photoionization cross sections, the recombination coefficient, K factors, Einstein A coefficients, and the two-photon decay rates (see~Refs.~\cite{Planck:2014ylh,Seto:2024cgo,Toda:2024ncp} for the detail).} as $E\propto m_{e}$. Since the energy level of hydrogen is higher with a larger electron mass, a photon with a temperature below standard recombination cannot excite it. Therefore, in the case that the electron mass in the recombination era is larger than today, the recombination occurs earlier than in the standard case as we show in Fig.~\ref{fig:rec}. Then, the earlier recombination leads to a higher Hubble constant $H_0$ and a smaller $\Omega_m$ to reproduce the measured CMB power spectrum as is observed.

\subsection{$w_0w_a$DE}
\label{subsec:w0wa} 

The dark energy model has two additional parameters:
the dark energy equation of state at the present time $w$ and its derivative with respect to the scale factor $w_a = dw/da$.
We adopt the so-called CPL parametrization~\cite{Chevallier:2000qy,Linder:2002et} of the equation of state
for a time-dependent dark energy as
\begin{align}
w(a)=w+w_a \left(1-\frac{a}{a_0}\right)=w+\frac{z}{1+z}w_a ,
\end{align}
where $a_0$ is the present value of scale factor $a$. In this paper, we only consider this parametrization, not other extended models of dark energy. It should be noted that decreasing $w$ leads to an increase in the Hubble constant $H_0$~\cite{Planck:2018vyg, Martinelli:2019krf, Vagnozzi:2019ezj, Alestas:2020mvb} and an increased $f\sigma_8$, as shown in Fig. 3 of Ref.~\cite{Tsujikawa:2012hv}.

\subsection{Extra radiation $N_{\mathrm{eff}}$}
Extra radiation model has one additional parameter: $N_{\mathrm{eff}}$ which is the effective number of neutrinos  defined as 
\begin{equation}
\Omega_{r}=\left(1+\frac{7}{8}\left(\frac{4}{11}\right)^{4/3}N_{\mathrm{eff}}\right)\Omega_{\gamma}.
\end{equation}
Here, $\gamma$ and $r$ denote CMB photons and radiation, respectively. We consider the extra radiation model to be a typical example of models where the energy density of other components is substantial during the early Universe, such as early dark energy~\cite{Poulin:2018dzj,Poulin:2018cxd,Agrawal:2019lmo,Smith:2019ihp,Lin:2019qug,Niedermann:2019olb,Braglia:2020bym,Niedermann:2020dwg,Ye:2020btb,Braglia:2020auw,Seto:2021xua, Rezazadeh:2022lsf}. We note that a large $\Delta N_\mathrm{eff}$ can be easily compatible with big bang nucleosynthesis if there is large lepton asymmetry~\cite{Seto:2021tad} or assuming the $\Delta N_{\mathrm{eff}}$ production after BBN~\cite{Allali:2024cji}.

As in the default setting of CAMB, in our calculation, we assume that the energy density of neutrinos is expressed by
\begin{align}
\Omega_{\nu}h^2 &= \frac{\sum m_{\nu}}{94.07\, \mathrm{eV}}\left(\frac{3.046}{3}\right)^{3/4}  \qquad (N_{\mathrm{eff}}>3.046),  \\
\Omega_{\nu}h^2 &= \frac{\sum m_{\nu}}{94.07\, \mathrm{eV}} \left( \frac{N_{\mathrm{eff}}}{3}\right)^{3/4} \qquad (\text{else}).
\end{align}
Such positive ($\Delta N_{\mathrm{eff}}>0$) is a case if additional massless particles besides neutrinos are produced, whereas negative ($\Delta N_{\mathrm{eff}}<0$) is realized by an additional entropy production after neutrino decoupling.

\section{datasets and methodology}
\label{sec:data}

We conduct a Markov-chain Monte Carlo (MCMC) analysis of four models using the public MCMC code \texttt{CosmoMC-planck2018}~\cite{Lewis:2002ah} to examine the effects of each model on a variety of cosmological observations with and without varying neutrino mass. We require the convergence $R-1<0.03$ and analyze the models by referring to the following cosmological observation.
We always used the CMB, BAO and light curve of SNeIa data, and we call this dataset $\mathcal{D}$.

\begin{itemize}
\item CMB from Planck~\cite{Planck:2018vyg}.
We use the temperature and polarization likelihoods for high $l$ \texttt{plik} ($l=30$ to $2508$ in TT and $l=30$
to $1997$ in EE and TE) and low$l$ \texttt{Commander} and lowE \texttt{SimAll}
($l=2$ to $29$). We also include CMB lensing~\cite{Planck:2018lbu}.
\item BAO distance data from 6dF~\cite{Beutler:2011hx}, MGS~\cite{Ross:2014qpa}, DR12~\cite{BOSS:2016wmc}, and DESI BAO~\cite{DESI:2024mwx}  (specific values are listed in Table~\ref{DESI-DATA}).
\item Light curve of SNeIa  from \textit{Pantheon}~\cite{Pan-STARRS1:2017jku}.
\item Priors on the values of $f \sigma_8(z)$ from the several observations~\cite{Avila:2021dqv, Said:2020epb, Simpson:2015yfa, Blake:2013nif, Blake:2011rj, Gil-Marin:2016wya, Mohammad:2018mdy, Song:2008qt, Okumura:2015lvp, eBOSS:2020gbb} (specific values are listed in Table~\ref{tab:fs8_table}). The covariance matrix of the RSD data is diagonal because the correlation coefficients are currently not available.  Hereafter, we call this ``RSD''.
\item A prior on the Hubble constant 
($H_0 = 73.30 \pm 1.04 $ km/sec/Mpc at 68\% CL) from the SH0ES Collaboration~\cite{Riess:2021jrx}. Hereafter, we call this ``R21''.
\end{itemize}

\begin{table}
\[
\begin{tabular}{|l|c|}
\hline
 \ensuremath{z_{\mathrm{eff}}}  &  \\
\hline\hline  6DF~\cite{Beutler:2011hx}&\\
 0.106  &  \ensuremath{r_{s}}/\ensuremath{D_{V}}=0.336\ensuremath{\pm}0.015\\ &  \\
\hline  MGS~\cite{Ross:2014qpa}&\\
 0.105  &  \ensuremath{D_{V}}/\ensuremath{r_{s}}=4.466\ensuremath{\pm}0.168\\ &  \\
\hline  DR12~\cite{BOSS:2016wmc}& \\
 0.38  &  \ensuremath{D_{M}}/\ensuremath{r_{s}}=1512.39\ensuremath{\pm}24.99\\
 0.38  &  H(z)\ensuremath{r_{s}}=81.2087\ensuremath{\pm}2.3683\\
 0.51  &  \ensuremath{D_{M}}/\ensuremath{r_{s}}=1975.22\ensuremath{\pm}30.10\\
 0.51  &  H(z)\ensuremath{r_{s}}=90.9029\ensuremath{\pm}2.3288\\
 0.61  &  \ensuremath{D_{M}}/\ensuremath{r_{s}}=2306.68\ensuremath{\pm}37.08\\
 0.61  &  H(z)\ensuremath{r_{s}}=98.9647\ensuremath{\pm} 2.5019\\ & \\
 \hline
\end{tabular}\;\;\;\;\;\begin{tabular}{|l|c|}
\hline
 \ensuremath{z_{\mathrm{eff}}}  &  \\
\hline\hline  DESI~\cite{DESI:2024mwx}&\\
 0.30  &  \ensuremath{D_{V}}/\ensuremath{r_{d}}=7.93\ensuremath{\pm}0.15\\
 0.51  &  \ensuremath{D_{M}}/\ensuremath{r_{d}}=13.62\ensuremath{\pm}0.25\\
 0.51  &  \ensuremath{D_{H}}/\ensuremath{r_{d}}=20.98\ensuremath{\pm}0.61\\
 0.71  &  \ensuremath{D_{M}}/\ensuremath{r_{d}}=16.85\ensuremath{\pm}0.32\\
 0.71  &  \ensuremath{D_{H}}/\ensuremath{r_{d}}=20.08\ensuremath{\pm}0.60\\
 0.93  &  \ensuremath{D_{M}}/\ensuremath{r_{d}}=21.71\ensuremath{\pm}0.28\\
 0.93  &  \ensuremath{D_{H}}/\ensuremath{r_{d}}=17.88\ensuremath{\pm}0.35\\
 1.32  &  \ensuremath{D_{M}}/\ensuremath{r_{d}}=27.79\ensuremath{\pm}0.69\\
 1.32  &  \ensuremath{D_{H}}/\ensuremath{r_{d}}=13.82\ensuremath{\pm}0.42\\
 1.49  &  \ensuremath{D_{V}}/\ensuremath{r_{d}}=26.07\ensuremath{\pm}0.67\\
 2.33  &  \ensuremath{D_{M}}/\ensuremath{r_{d}}=39.71\ensuremath{\pm}0.94\\
 2.33  &  \ensuremath{D_{H}}/\ensuremath{r_{d}}=8.52\ensuremath{\pm}0.17 \\ & \\
 \hline
\end{tabular}
\]

\caption{The distance and expansion rete data from 6DF, MGS, and DR12 (left).
The distant data from DESI (right). \label{DESI-DATA}}
\end{table}

\begin{table}[t!]
\begin{center}
\begin{tabular}{| c | c |c | c |}
\multicolumn{1}{c}{Survey} &  \multicolumn{1}{c}{$z$} &  \multicolumn{1}{c}{$f(z)\sigma_8(z)$} & \multicolumn{1}{c}{{\small References}}
\\\hline
ALFALFA & $0.013$ & $0.46\pm 0.06$ & \cite{Avila:2021dqv}
\\\hline
6dFGS+SDSS & $0.035$ & $0.338\pm 0.027$ & \cite{Said:2020epb}
\\\hline
GAMA & $0.18$ & $0.29\pm 0.10$ & \cite{Simpson:2015yfa}
\\ \cline{2-4}& $0.38$ & $0.44\pm0.06$ & \cite{Blake:2013nif}
\\\hline
 WiggleZ & $0.22$ & $0.42\pm 0.07$ & \cite{Blake:2011rj} \tabularnewline
\cline{2-3} & $0.41$ & $0.45\pm0.04$ & \tabularnewline
\cline{2-3} & $0.60$ & $0.43\pm0.04$ & \tabularnewline
\cline{2-3} & $0.78$ & $0.38\pm0.04$ &
\\\hline
DR12 BOSS & $0.32$ & $0.427\pm 0.056$  & \cite{Gil-Marin:2016wya}\\ \cline{2-3}
 & $0.57$ & $0.426\pm 0.029$ &
\\\hline
VIPERS & $0.60$ & $0.49\pm 0.12$ & \cite{Mohammad:2018mdy}
\\ \cline{2-3}& $0.86$ & $0.46\pm0.09$ &
\\\hline
VVDS & $0.77$ & $0.49\pm0.18$ & \cite{Song:2008qt}
\\\hline
FastSound & $1.36$ & $0.482\pm0.116$ & \cite{Okumura:2015lvp}
\\\hline
eBOSS Quasar & $1.48$ & $0.462\pm 0.045$ & \cite{eBOSS:2020gbb}
\\\hline
 \end{tabular}
\end{center}
\caption{Specific values of $f(z)\sigma_8(z)$ and their errors for each measurements.}
\label{tab:fs8_table}
\end{table}

\section{Results}
\label{sec:results}

In the Table~\ref{Tab:68}, we summarize the constraints of the cosmological parameters and the Gaussian tension which is defined as,
\begin{equation}  T_{H_0}=\frac{H_{0~\mathcal{D}+\mathrm{RSD}}-73.
30\, \mathrm{km/s/Mpc}}{\sqrt{\sigma_{\mathcal{D+\mathrm{RSD}}}^{2}+(1.04\, \mathrm{km/s/Mpc})}^{2}} \,,
\end{equation}
for the Hubble tension with the direct measurement of the Hubble constant~\cite{Riess:2021jrx} and
\begin{equation}
T_{S_8}=\frac{S_{8~\mathcal{D}+\mathrm{RSD}}-0.776}{\sqrt{\sigma_{\mathcal{D}+\mathrm{RSD}}^{2}+\frac{0.017^{2}+0.017^{2}}{2}}} \,,
\end{equation} 
for the $S_{8}$ tension with DES~\cite{DES:2021wwk}. 
We define $S_8=\sigma_8 \sqrt{\Omega_m/0.3}$.

In the Table~\ref{Tab:best}, we summarize the best-fit values of the cosmological parameters and their corresponding minimized chi-squared values. To penalize increasing the number of parameters and facilitate fair comparison, we also calculate the Akaike information criterium (AIC) of model $X$ relative to that of $\Lambda$CDM model as follows:
\begin{equation}
    \Delta AIC = \chi^2_{\mathrm{min}, X} -\chi^2_{\mathrm{min},\Lambda\mathrm{CDM}}
    + 2(N_{X}-N_{\Lambda \mathrm{CDM}}),
\end{equation}
where $N_{X}-N_{\Lambda \mathrm{CDM}}$ is the number difference in the free parameters between the model $X$ and $\Lambda$CDM model.

\begin{table}[h]
\[
\begin{tabular}{lcccc}
 Parameter  &  LCDM  &  LCDM+\ensuremath{\sum m_{\nu}}  &  me  &  me+\ensuremath{\sum m_{\nu}}\\
\hline  {\boldmath\ensuremath{\Omega_{b}h^{2}}}  &  \ensuremath{0.02255\pm0.00013}  &  \ensuremath{0.02253\pm0.00013}  &  \ensuremath{0.02261\pm0.00014}  &  \ensuremath{0.02271_{-0.00022}^{+0.00017}}\\
 {\boldmath\ensuremath{\Omega_{c}h^{2}}}  &  \ensuremath{0.11770\pm0.00074}  &  \ensuremath{0.11784\pm0.00076}  &  \ensuremath{0.1194\pm0.0017}  &  \ensuremath{0.1206_{-0.0026}^{+0.0020}}\\
 {\boldmath\ensuremath{\Sigma m_{\nu}}}  &    &  \ensuremath{<0.111}  &    &  \ensuremath{<0.330}\\
 {\boldmath\ensuremath{m_{e}/m_{e0}}}  &    &    &  \ensuremath{1.0058\pm0.0053}  &  \ensuremath{1.0114_{-0.011}^{+0.0069}}\\
 \ensuremath{H_{0}}  &  \ensuremath{68.41\pm0.34}  &  \ensuremath{68.49\pm0.38}  &  \ensuremath{69.23\pm0.83}  &  \ensuremath{69.66_{-1.1}^{+0.93}}\\
 \ensuremath{\Omega_{m}}  &  \ensuremath{0.3011\pm0.0043}  &  \ensuremath{0.3003\pm0.0047}  &  \ensuremath{0.2978\pm0.0053}  &  \ensuremath{0.2982\pm0.0054}\\
 \ensuremath{S_{8}}  &  \ensuremath{0.8058\pm0.0085}  &  \ensuremath{0.8079\pm0.0094}  &  \ensuremath{0.8087\pm0.0090}  &  \ensuremath{0.804_{-0.010}^{+0.013}}\\
 \ensuremath{r_{{\rm drag}}h}  &  \ensuremath{100.90\pm0.57}  &  \ensuremath{101.01\pm0.62}  &  \ensuremath{101.38\pm0.73}  &  \ensuremath{101.36\pm0.73}\\
 \ensuremath{f\sigma_{8}(0.38)}  &  \ensuremath{0.4672\pm0.0040}  &  \ensuremath{0.4685_{-0.0043}^{+0.0048}}  &  \ensuremath{0.4700\pm0.0048}  &  \ensuremath{0.4681_{-0.0053}^{+0.0064}}\\
 \ensuremath{f\sigma_{8}(0.61)}  &  \ensuremath{0.4630\pm0.0035}  &  \ensuremath{0.4643_{-0.0040}^{+0.0046}}  &  \ensuremath{0.4664\pm0.0048}  &  \ensuremath{0.4645_{-0.0054}^{+0.0064}}\\
\hline  \ensuremath{T_{H_{0}}} & 4.46\ensuremath{\sigma} & 4.34\ensuremath{\sigma} & 3.05\ensuremath{\sigma} & 2.50\ensuremath{\sigma}\\
 \ensuremath{T_{S_{8}}}  & 1.56\ensuremath{\sigma} & 1.64\ensuremath{\sigma} & 1.70\ensuremath{\sigma} & 1.36\ensuremath{\sigma}
\end{tabular}
\]
\[
\begin{tabular}{lcccc}
 Parameter  &  \ensuremath{w_{0}w_{a}} CDM  &  \ensuremath{w_{0}w_{a}} CDM+\ensuremath{\sum m_{\nu}}  &  \ensuremath{N_{\mathrm{eff}}}  &  \ensuremath{N_{\mathrm{eff}}}+\ensuremath{\sum m_{\nu}}\\
\hline  {\boldmath\ensuremath{\Omega_{b}h^{2}}}  &  \ensuremath{0.02249\pm0.00014}  &  \ensuremath{0.02249\pm0.00013}  &  \ensuremath{0.02249\pm0.00017}  &  \ensuremath{0.02246\pm0.00018}\\
 {\boldmath\ensuremath{\Omega_{c}h^{2}}}  &  \ensuremath{0.11841\pm0.00094}  &  \ensuremath{0.11835\pm0.00094}  &  \ensuremath{0.1165\pm0.0027}  &  \ensuremath{0.1160\pm0.0027}\\
 {\boldmath\ensuremath{\Sigma m_{\nu}}}  &   &  \ensuremath{<0.195}  &   &  \ensuremath{<0.103}\\
 {\boldmath\ensuremath{w_{0}}},{\boldmath\ensuremath{N_{\mathrm{eff}}}}  &  \ensuremath{-0.900\pm0.069}  &  \ensuremath{-0.894\pm0.067}  &  \ensuremath{2.97\pm0.16}  &  \ensuremath{2.93\pm0.17}\\
 {\boldmath\ensuremath{w_{a}}}  &  \ensuremath{-0.38\pm0.24}  &  \ensuremath{-0.41_{-0.24}^{+0.28}}\\
 \ensuremath{H_{0}}  &  \ensuremath{68.04\pm0.72}  &  \ensuremath{68.00\pm0.74}  &  \ensuremath{67.9\pm1.0}  &  \ensuremath{67.9\pm1.0}\\
 \ensuremath{\Omega_{m}}  &  \ensuremath{0.3058\pm0.0069}  &  \ensuremath{0.3066\pm0.0071}  &  \ensuremath{0.3025\pm0.0050}  &  \ensuremath{0.3017\pm0.0050}\\
 \ensuremath{S_{8}}  &  \ensuremath{0.8135\pm0.0099}  &  \ensuremath{0.812\pm0.011}  &  \ensuremath{0.8037\pm0.0094}  &  \ensuremath{0.8057\pm0.0098}\\
 \ensuremath{r_{{\rm drag}}h}  &  \ensuremath{100.3\pm1.1}  &  \ensuremath{100.2\pm1.1}  &  \ensuremath{100.75\pm0.63}  &  \ensuremath{100.88\pm0.64}\\
 \ensuremath{f\sigma_{8}(0.38)}  &  \ensuremath{0.4660\pm0.0065}  &  \ensuremath{0.4655\pm0.0068}  &  \ensuremath{0.4656\pm0.0050}  &  \ensuremath{0.4667\pm0.0052}\\
 \ensuremath{f\sigma_{8}(0.61)}  &  \ensuremath{0.4636\pm0.0069}  &  \ensuremath{0.4632\pm0.0072}  &  \ensuremath{0.4611\pm0.0050}  &  \ensuremath{0.4623\pm0.0052}\\
\hline  \ensuremath{T_{H_{0}}} & 4.16\ensuremath{\sigma} & 4.15\ensuremath{\sigma} & 3.74\ensuremath{\sigma} & 3.74\ensuremath{\sigma}\\
 \ensuremath{T_{S_{8}}}  & 1.91\ensuremath{\sigma} & 1.78\ensuremath{\sigma} & 1.43\ensuremath{\sigma} & 1.51\ensuremath{\sigma}
\end{tabular}
\]

\caption{95\% upper limit of neutrino mass and 68\% C.L of the other parameters. We use CMB, Pantheon, BAO distance and RSD data.  \label{Tab:68}}
\end{table}

\begin{sidewaystable}[h]
\[
\begin{tabular}{lcccccccc}
 Parameter  &  LCDM  &  LCDM+\ensuremath{\sum m_{\nu}}  &  \ensuremath{w_{0}w_{a}}DE  &  \ensuremath{w_{0}w_{a}}DE+\ensuremath{\sum m_{\nu}}  &  \ensuremath{N_{\mathrm{eff}}}  &  \ensuremath{N_{\mathrm{eff}}}+\ensuremath{\sum m_{\nu}}  &  \ensuremath{m_{e}}  &  \ensuremath{m_{e}}+\ensuremath{\sum m_{\nu}} \\
\hline  {\boldmath\ensuremath{\Sigma m_{\nu}}}  &  -  &  0.0519  &  -  &  0.0684  &  -  &  0.0434  &  -  &  0.0711\\
 {\boldmath\ensuremath{w_{0}}} ,{\boldmath\ensuremath{N_{\mathrm{eff}}}},{\boldmath\ensuremath{m_{e}/m_{e0}}}  &  -  &  -  &  -1.0041  &  -1.0109  &  3.217  &  3.254  &  1.018  &  1.021\\
 {\boldmath\ensuremath{w_{a}}}  &  -  &  -  &  -0.0113  &  -0.0115  &  -  &  -  &  -  &  -\\
 \ensuremath{H_{0}}  &  68.90  &  68.93  &  69.03  &  69.09  &  69.73  &  70.08  &  71.28  &  71.62\\
 \ensuremath{\Omega_{m}}  &  0.2951  &  0.2949  &  0.2943  &  0.2945  &  0.2945  &  0.2927  &  0.2871  &  0.2865\\
 \ensuremath{S_{8}}  &  0.7975  &  0.8003  &  0.7987  &  0.7994  &  0.8031  &  0.8068  &  0.8165  &  0.8150\\
\hline  \ensuremath{\chi_{\mathrm{CMB}}^{2}}  &  2780.696  &  2780.445  &  2779.891  &  2778.858  &  2780.831  &  2780.430  &  2775.648  &  2776.177\\
 \ensuremath{\chi_{\mathrm{Pantheon}}^{2}}  &  1034.765  &  1034.767  &  1034.822  &  1034.892  &  1034.775  &  1034.816  &  1035.033  &  1035.062\\
 \ensuremath{\chi_{\mathrm{RSD}}^{2}}  &  22.572  &  23.222  &  23.117  &  23.609  &  23.975  &  24.970  &  28.609  &  28.276\\
 \ensuremath{\chi_{\mathrm{BAO\,distance}}^{2}}  &  19.179  &  19.189  &  19.424  &  19.530  &  19.222  &  19.613  &  23.234  &  23.950\\
 \ensuremath{\chi_{H_{0}}^{2}}  &  17.903  &  17.666  &  16.847  &  16.387  &  11.778  &  9.594  &  3.759  &  2.619\\
 \ensuremath{\chi_{\mathrm{prior}}^{2}}  &  2.304  &  1.502  &  1.842  &  1.895  &  2.048  &  1.830  &  1.791  &  1.731\\
\hline  \ensuremath{\chi_{\mathrm{total}}^{2}}  &  3877.42  &  3876.79  &  3875.94  &  3875.17  &  3872.63  &  3871.25  &  3868.07  &  3867.82\\
 \ensuremath{\chi_{\mathrm{total}}^{2}}-\ensuremath{\chi_{\mathrm{total\,\Lambda CDM}}^{2}}  &   &  -0.63  &  -1.48  &  -2.25  &  -4.79  &  -6.17  &  -9.35  &  -9.6\\
 \ensuremath{\Delta}AIC  &   &  1.37  &  2.52  &  3.75  &  -2.79  &  -2.17  &  -7.35  &  -5.6\\
 
\end{tabular}
\]

\caption{The best-fit values of the cosmological parameters and their corresponding minimized chi-squared values. We use the data$\mathcal{D}$+RSD+R21. \label{Tab:best}}
\end{sidewaystable}

\begin{figure}[ht]
\includegraphics[width=16.2cm]{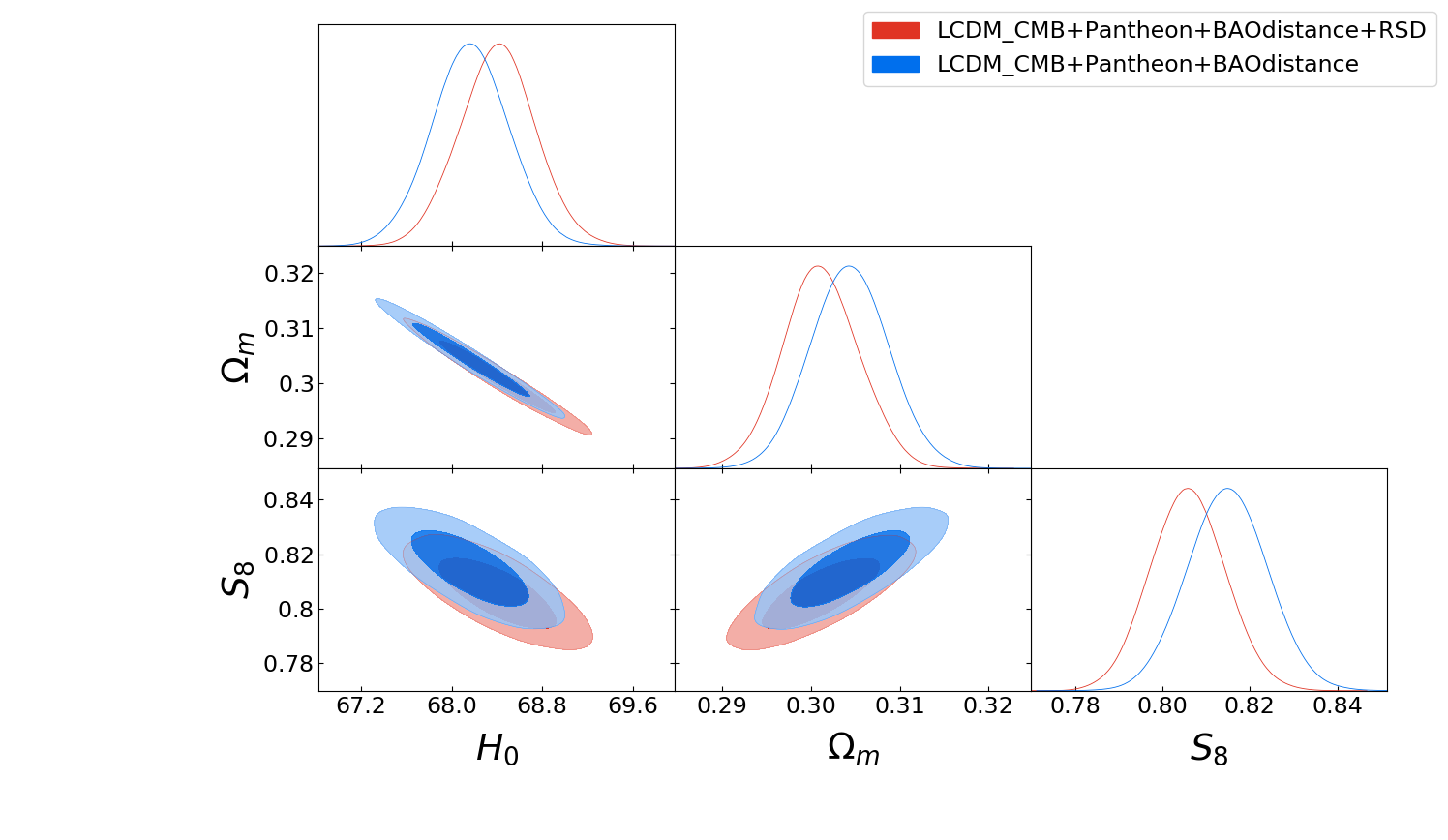}
\includegraphics[width=16.2cm]{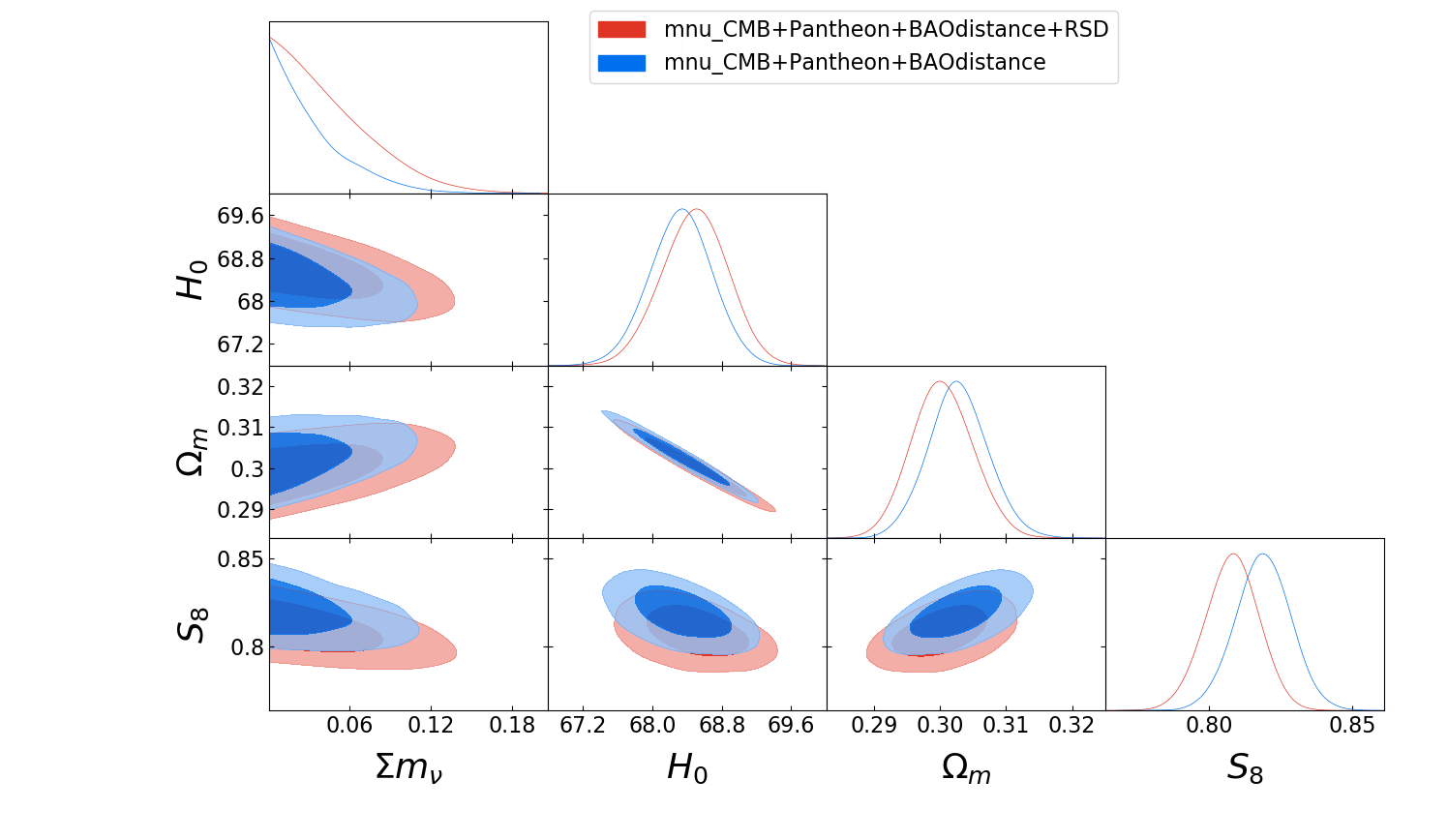}
\caption{Posterior distributions of the several parameters on $\Lambda$CDM model and $\Lambda$CDM+$m_{\nu}$ model for the data $\mathcal{D}$ with/without RSD.}
\label{fig:LCDM} 
\end{figure}

\begin{figure}[ht]
\includegraphics[width=16cm]{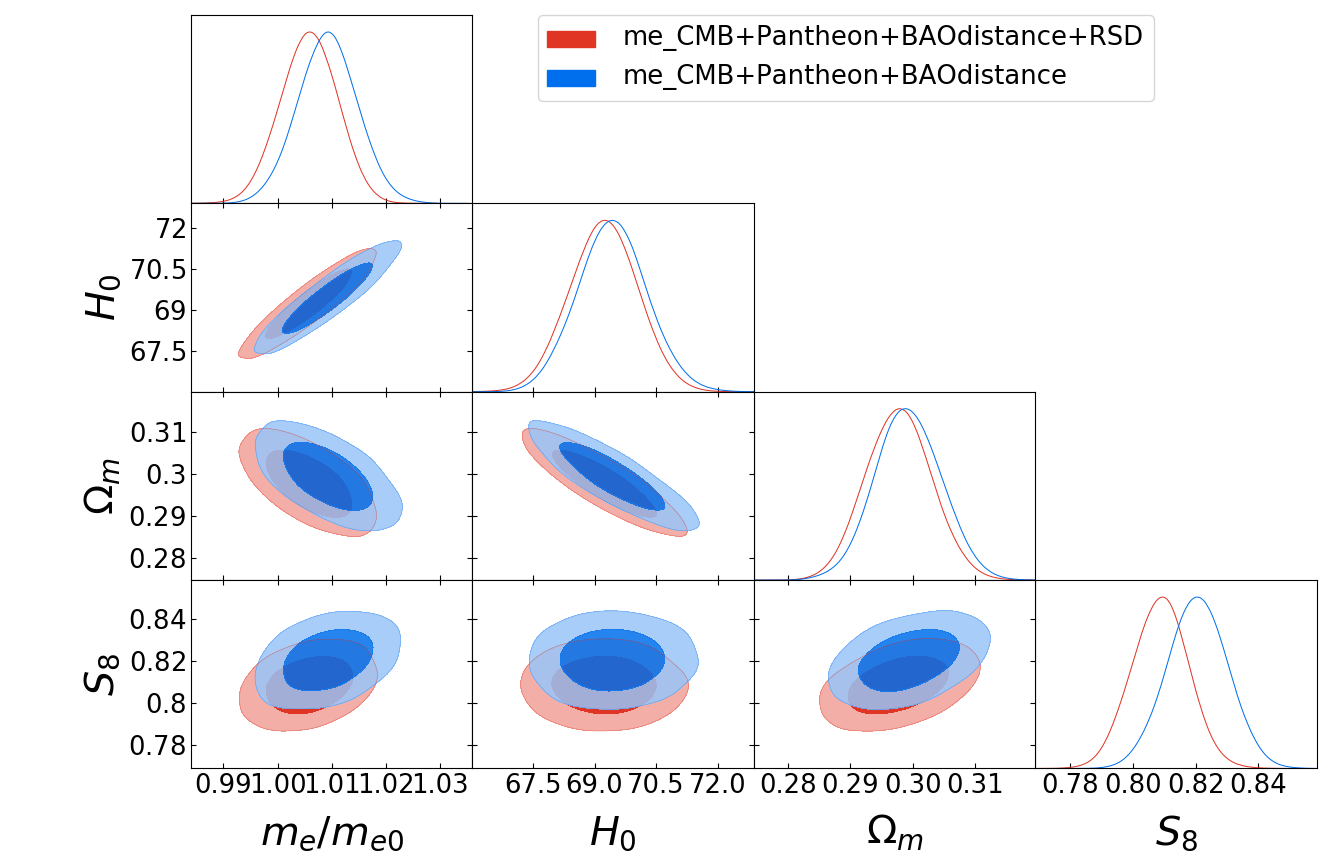}
\includegraphics[width=16cm]{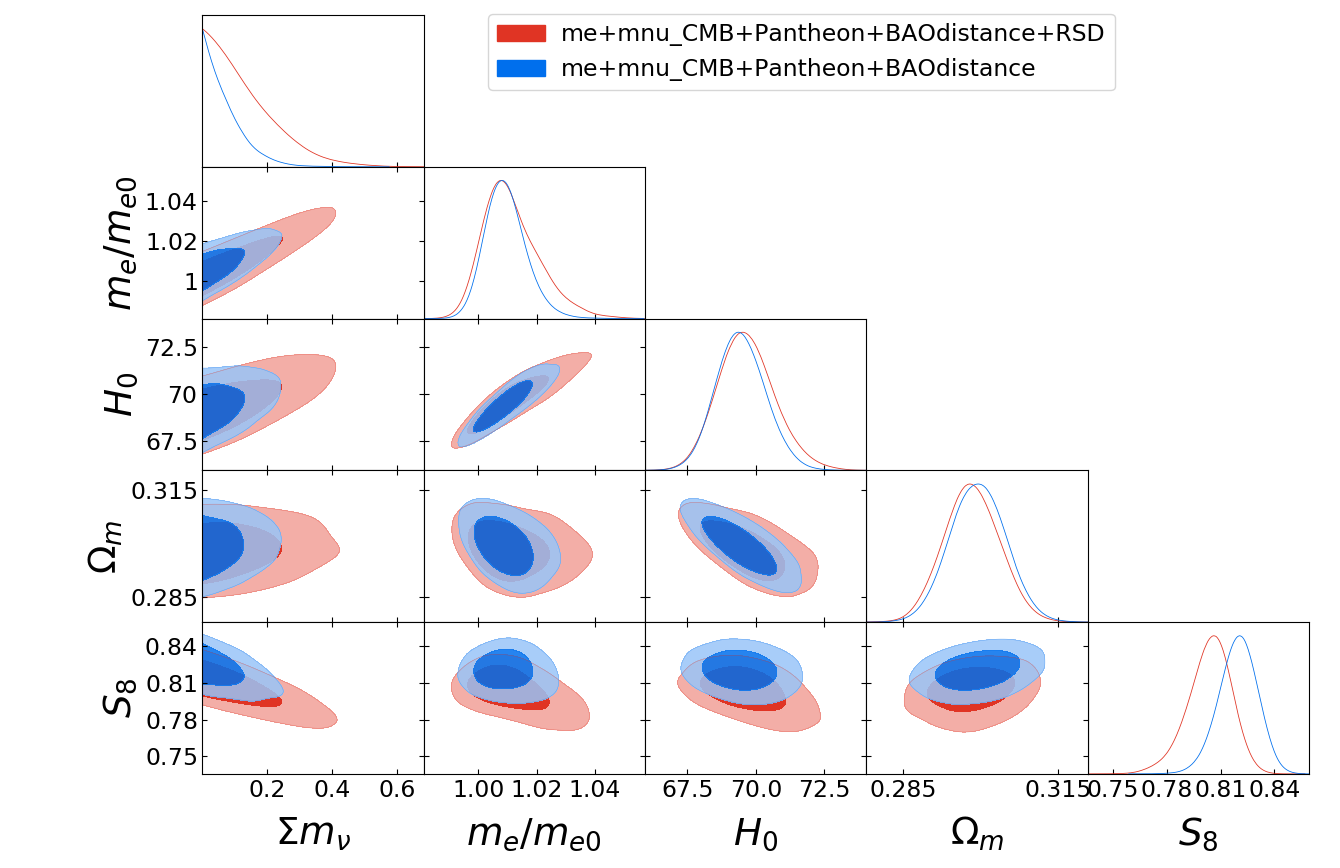}
\caption{Posterior distributions of the several parameters on varying $m_e$ model and varying $m_e$+$m_{\nu}$ model for the data $\mathcal{D}$ with/without RSD.}
\label{fig:me} 
\end{figure}

\begin{figure}[ht]
\includegraphics[width=16cm]{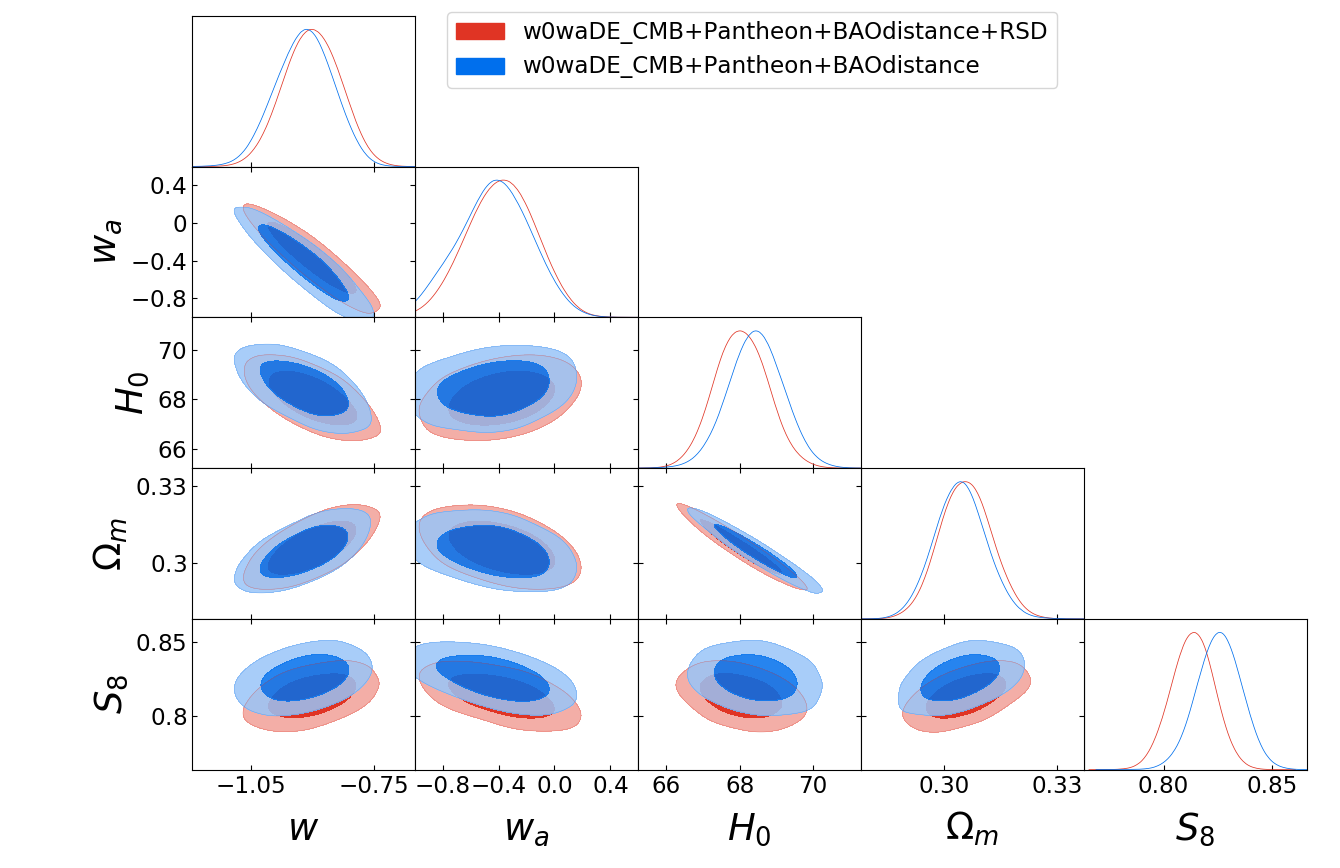}
\includegraphics[width=16cm]{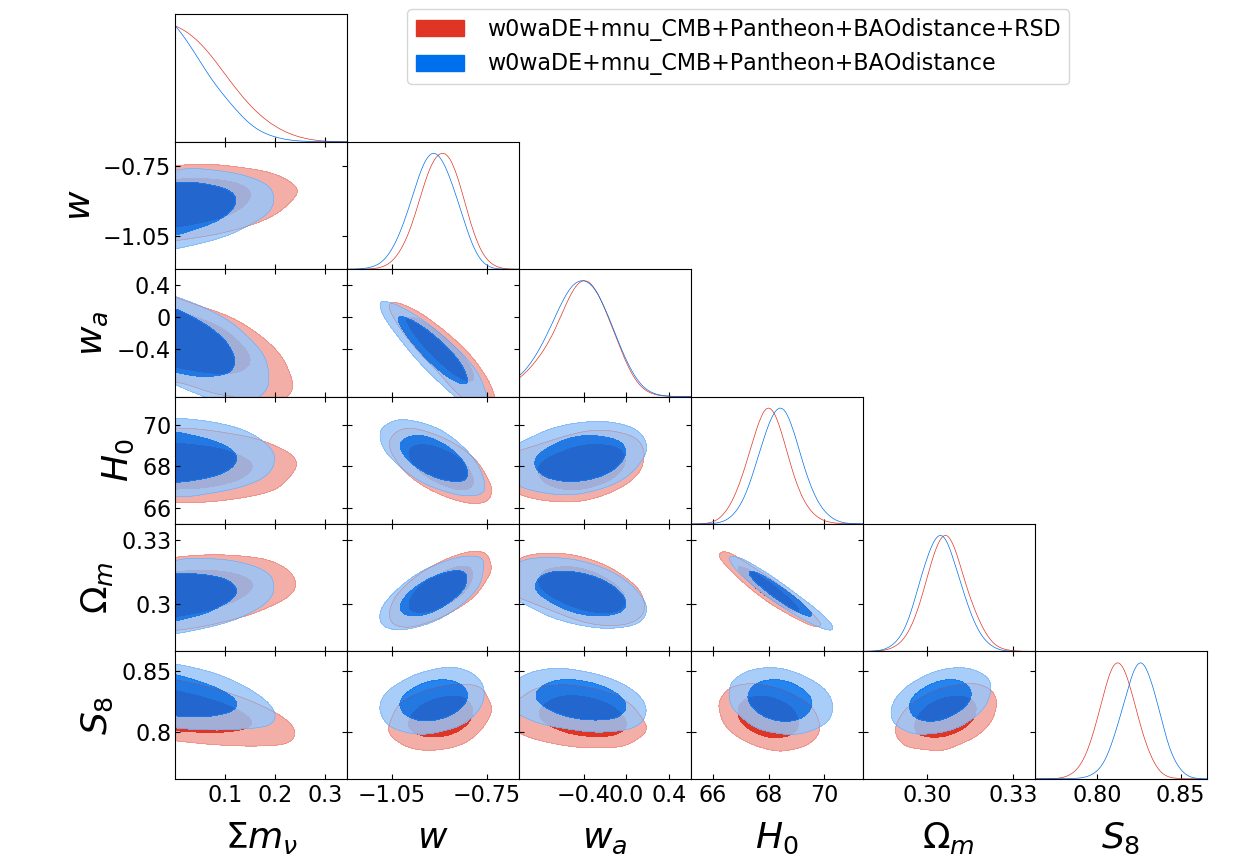}
\caption{Posterior distributions of the several parameters on $w_0w_a$DE model and $w_0w_a$DE+$m_{\nu}$ model for the data $\mathcal{D}$ with/without RSD.}
\label{fig:w0waDE} 
\end{figure}

\begin{figure}[ht]
\includegraphics[width=16cm]{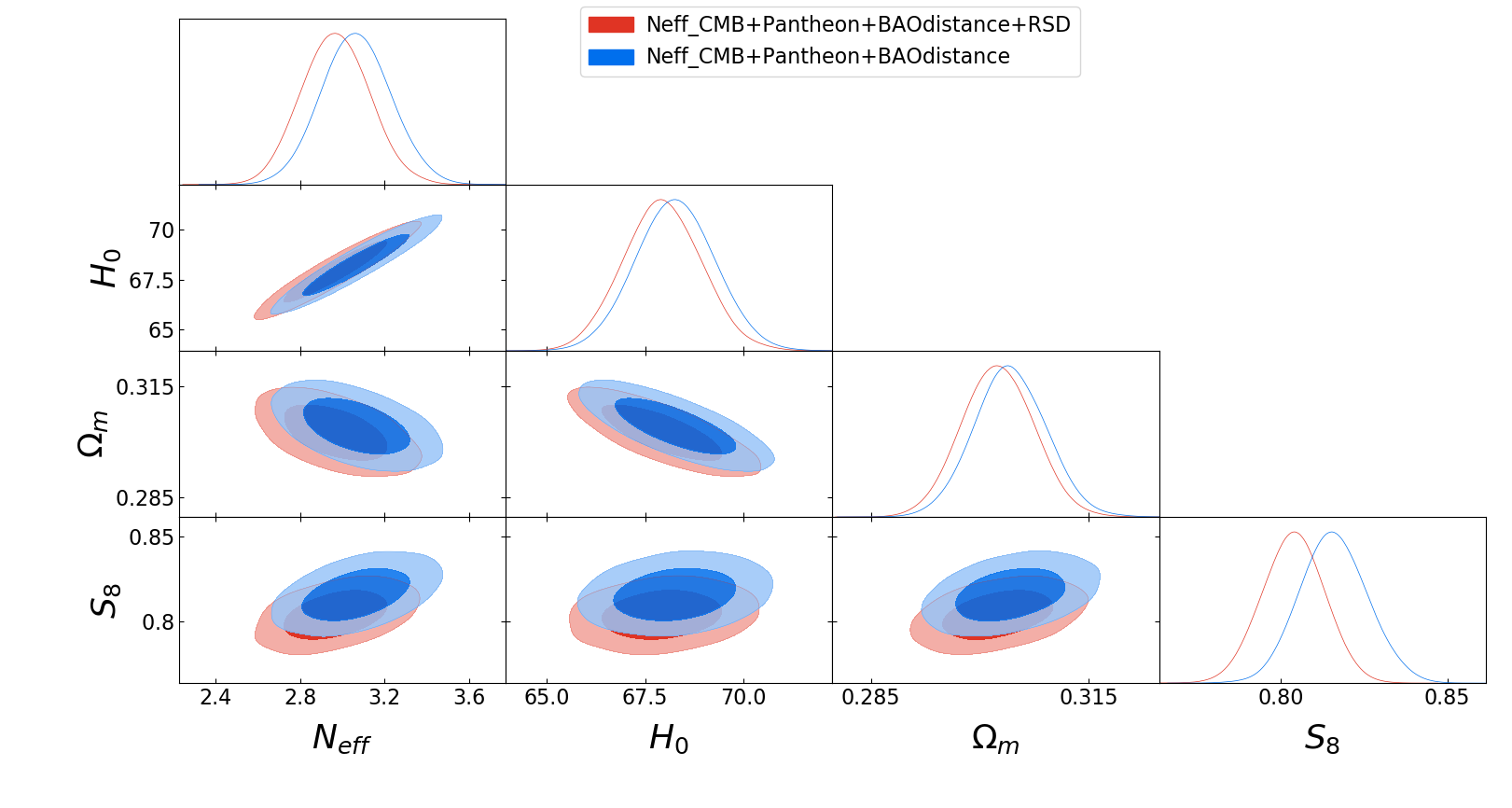}
\includegraphics[width=16cm]{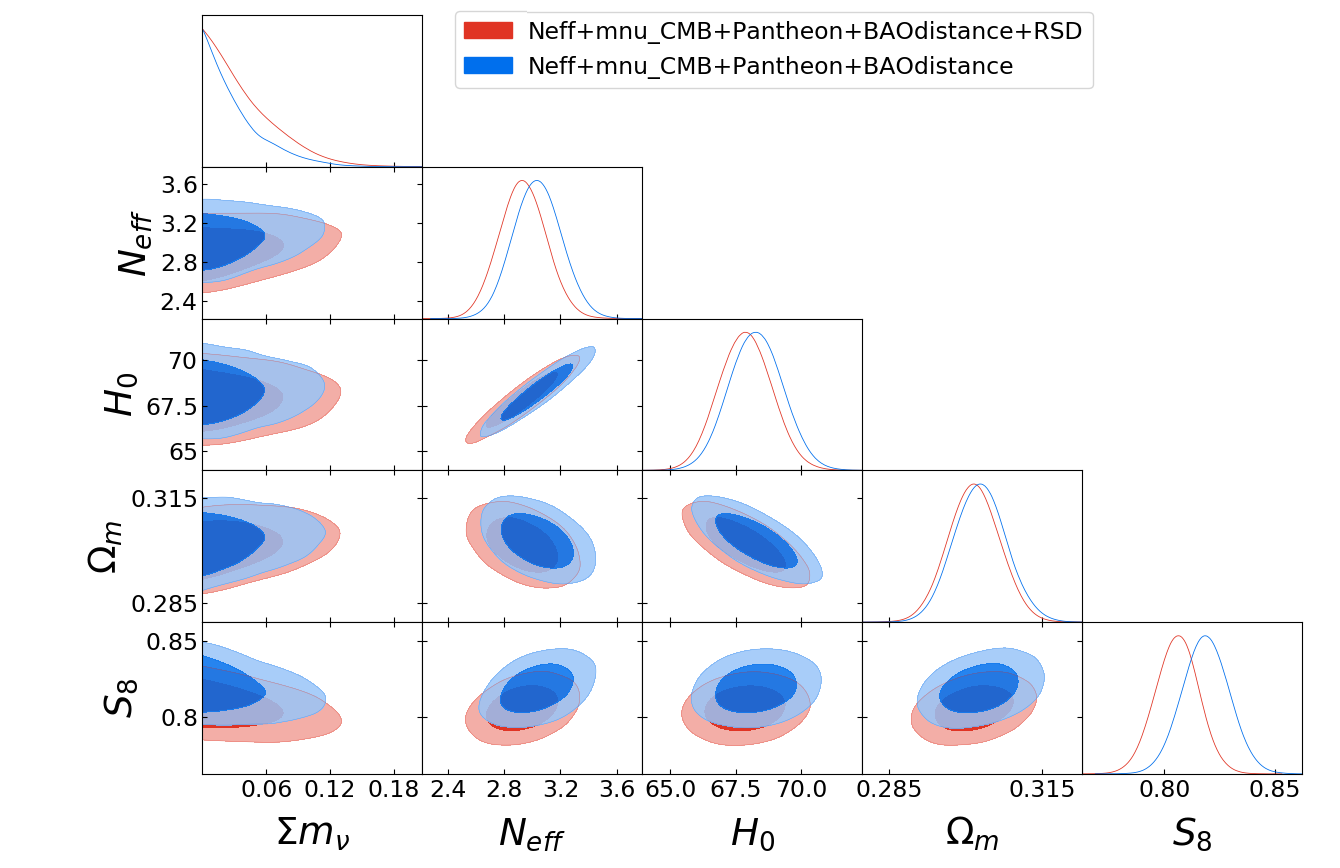}
\caption{Posterior distributions of the several parameters on $N_{\mathrm{eff}}$ model and $N_{\mathrm{eff}}$+$m_{\nu}$ model for the data $\mathcal{D}$ with/without RSD.}
\label{fig:Neff} 
\end{figure}

\begin{figure}[ht]
\includegraphics[width=16cm]{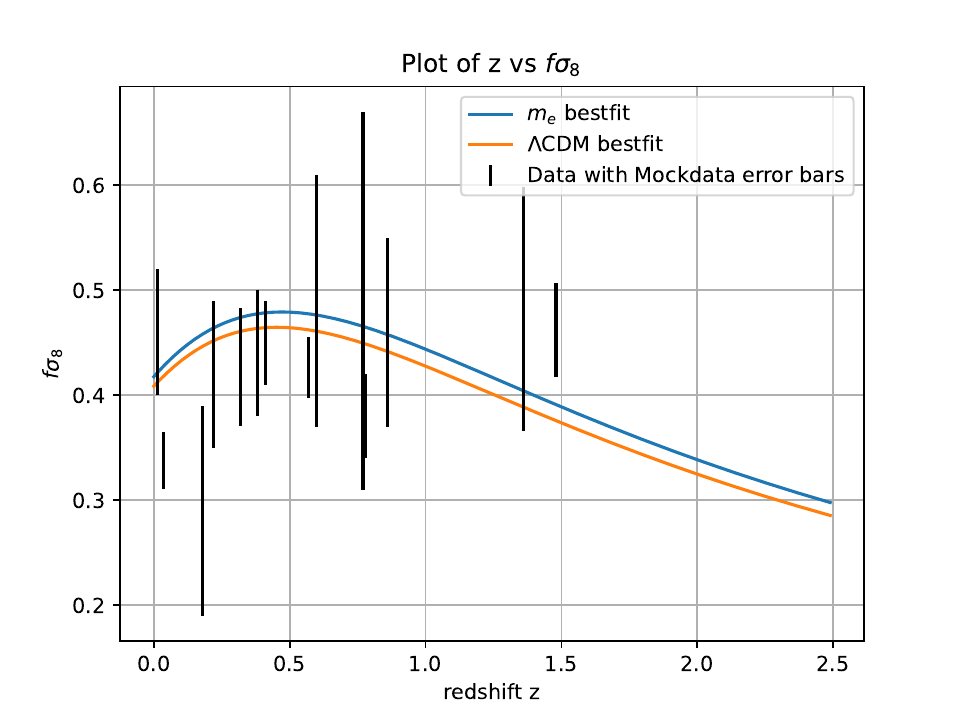}
\caption{The plot of $f\sigma_8(z)$ for the best-fit value of $\Lambda$CDM model $\left[ (\Omega_m, \sigma_8)=(0.2951,0.8041) \right]$ and varying electron mass model $\left[ (\Omega_m, \sigma_8)=(0.2871,0.8347) \right]$.}
\label{fig:fsigma} 
\end{figure}

\subsection{$\Lambda$CDM}
The 68\% constraints and best-fit values of the cosmological parameters on $\Lambda$CDM model and the $\Lambda$CDM + $m_\nu$ model are summarized in Table~\ref{Tab:68} and Table~\ref{Tab:best}. The one-dimensional posterior distributions and two-dimensional allowed regions are shown in Fig.~\ref{fig:LCDM}.

The inclusion of RSD data results in the preference of a higher Hubble constant and lower $S_8$ , which alleviate both the Hubble tension and the $S_8$ tension,  for the simple $\Lambda$CDM (fixed neutrino mass) model as well as to allow a larger neutrino mass for the $\Lambda$CDM + $m_\nu$ model, as we discussed in Sec.~\ref{subsec:mnu}. We obtain the 95\% upper limits of neutrino mass as,
\begin{align}
\sum m_\nu&<0.0880~\mathrm{eV} \qquad  (\mathcal{D}), \label{eq:5} \\
\sum m_\nu&<0.111~\mathrm{eV}  \qquad (\mathcal{D}+\mathrm{RSD}). \label{eq:6}
\end{align}
Both inequalities~(\ref{eq:5}) and (\ref{eq:6}) are consistent with the fact that the sum of neutrino mass is larger than $0.06~\mathrm{eV}$ from observations of  neutrino oscillation.

\subsection{Varying $m_e$}

The 68\% constraints and best-fit values of the cosmological parameters on the varying $m_e$ model and the varying $m_e$ + $m_\nu$ model are summarized in Table~\ref{Tab:68} and Table~\ref{Tab:best}. The-one dimensional posterior distributions and two-dimensional allowed regions are shown in Fig.~\ref{fig:me}.

For the varying $m_e$ (fixed neutrino mass) model, when we include RSD data, we find that a smaller electron mass and a slightly lower Hubble constant are preferred as, 
\begin{align}
&m_e/m_{e0}=1.0092\pm0.0055,  \notag \\
&H_0=69.44\pm0.84~\mathrm{km/s/Mpc}\,\,\,\,(\mathcal{D}), 
\end{align}
and
\begin{align}
&m_e/m_{e0}=1.0058\pm0.0053, \notag \\
&H_0=69.23\pm0.83~\mathrm{km/s/Mpc} \qquad  (\mathcal{D}+\mathrm{RSD}).
\end{align}
It is understood, as follows, that a larger electron mass $m_e/m_{e0}$ also leads to too large $\sigma_8$ and too large growth rate $f\sigma_8(z)$ to fit with the RSD data.
This can also be seen by the Table~\ref{Tab:best} and Fig.~\ref{fig:fsigma}. 
In the Table~\ref{Tab:best}, 
though the varying $m_e$ model greatly improves the total chi-squared values,
the chi-squared value of RSD becomes worse from $\chi^2_{\mathrm{RSD}}=22.572$ in $\Lambda$CDM model to $\chi^2_{\mathrm{RSD}}=28.609$ in varying $m_e$ model.
In the Fig.~\ref{fig:fsigma}, we show a growth rates $f\sigma_8(z)$ as a function of redshift, and we draw the blue curve using the best-fit values of the varying $m_e$ model and the orange curve using the best-fit values of $\Lambda$CDM model.

On the other hand, a larger neutrino mass is allowed in the varying $m_e$ + $m_\nu$ model when we include the RSD data.
It is fascinating to observe that a larger neutrino mass leads to a larger electron mass, which in turn leads to a higher Hubble constant.
Thus, we obtain the 68\% constraints of $m_e/m_{e0}$ and $H_0$ and the 95\% upper limits of neutrino mass as,
\begin{align}
&m_e/m_{e0}=1.0094^{+0.0059}_{-0.0077},  \notag \\
&H_0=69.45^{+0.85}_{-0.95}~\mathrm{km/s/Mpc}, \notag \\
&\sum m_\nu<0.193~\mathrm{eV}\qquad(\mathcal{D}), 
\end{align}
\begin{align}
&m_e/m_{e0}=1.0114^{+0.0069}_{-0.011}, \notag \\
&H_0=69.66^{+0.93}_{-1.1}~\mathrm{km/s/Mpc}, \notag \\
&\sum m_\nu<0.330~\mathrm{eV}\qquad  (\mathcal{D}+\mathrm{RSD}).
\end{align}
It is worth noting that the varying $m_e$ + $m_\nu$ model is the best solution to both the Hubble and $S_8$ tension as far as we see in Table~\ref{Tab:68}.

\subsection{$w_0w_a$DE}
The 68\% constraints and best-fit values of the cosmological parameters in the $w_0w_a$DE model and $w_0w_a$DE + $m_\nu$ model 
 are summarized in Table~\ref{Tab:68} and Table~\ref{Tab:best}. 
 The-one dimensional posterior distributions and two-dimensional allowed regions are illustrated in Fig.~\ref{fig:w0waDE}.

For the varying $w_0w_a$DE (fixed neutrino mass) model, the recent DESI paper~\cite{DESI:2024mwx} discovered that the dynamical ($w>-1$) dark energy is preferred at a significance level of approximately 2.6$\sigma$ using CMB(Planck and Atacama Cosmology Telescope data), Pantheon+~\cite{Brout:2022vxf} and DESI BAO data. We also find the preference of the dynamical dark energy as,
\begin{align}
&w=-0.919\pm0.069, \notag \\
&w_a=-0.42\pm0.25, \notag \\
&H_0=68.44\pm0.75~\mathrm{km/s/Mpc}\qquad(\mathcal{D}), 
\end{align}
\begin{align}
&w=-0.900\pm0.069, \notag \\
&w_a=-0.38\pm0.24, \notag \\
&H_0=68.04\pm0.72~\mathrm{km/s/Mpc}\qquad  (\mathcal{D}+\mathrm{RSD}).
\end{align}
Including the RSD data results in a preference for larger $w$ and $w_a$ because a larger $w$ results in a smaller $f\sigma_8$ and is consistent with the observed RSD. By having a larger $w$, the Hubble constant decreases, which takes the Hubble tension more seriously.
In the Table~\ref{Tab:68}, we show that the center values of the Hubble constant get lower than the $\Lambda$CDM case, although the tensions $T_{H_0}$ are slightly improved due to larger errors.

For the varying $w_0w_a$DE + $m_\nu$ model, it is worth mentioning that a larger neutrino mass is allowed than $\Lambda$CDM case (Eqs.~(\ref{eq:5}) and (\ref{eq:6})) and its 95\% upper limits are given by
\begin{align}
\sum m_\nu&<0.159~\mathrm{eV} \qquad  (\mathcal{D}),  \\
\sum m_\nu&<0.195~\mathrm{eV} \qquad  (\mathcal{D}+\mathrm{RSD}). 
\end{align}

\subsection{Extra radiation $N_{\mathrm{eff}}$}

The 68\% constraints and best-fit values of the cosmological parameters on extra radiation $N_{\mathrm{eff}}$ model and extra radiation $N_{\mathrm{eff}}$ + $m_\nu$ model 
 are summarized in Table~\ref{Tab:68} and Table~\ref{Tab:best}. 
 The-one dimensional posterior distributions and two-dimensional allowed regions are illustrated in Fig.~\ref{fig:Neff}.

For both the $N_\mathrm{eff}$ model and $N_\mathrm{eff}$ + $m_\nu$ model, when we include RSD data, we find that a smaller $N_{\mathrm{eff}}$, a lower Hubble constant $H_0$, and a lower $S_8$ are preferred.
As in Table~\ref{Tab:best}, in the $N_\mathrm{eff}$ model and  $N_\mathrm{eff}$ + $m_\nu$ model, the best-fit values of the Hubble constant are around $70~\mathrm{km/s/Mpc}$, and they do not get the value of $\chi^2_{\mathrm{RSD}}$ so worse as the varying $m_e$ model does.

\section{Conclusions}
\label{sec:conclusions}

In this paper, we examine several cosmological models and how they affect observations, with particular emphasis on RSD. 

The analysis with RSD data reveals that $S_8$ and $f\sigma_8$ are smaller, which can be seen in the results for $\Lambda$CDM in Fig.~\ref{fig:LCDM}. Adding RSD allows for additional parameters and effects to reduce $\sigma_8$, including the neutrinos mass. In fact, when we include the RSD data into analysis, we found a slightly weaker $95\%$ upper limit $\sum m_\nu<0.111~\mathrm{eV} (\mathcal{D}+\mathrm{RSD})$ than the $95\%$ upper limit $\sum m_\nu<0.0880~\mathrm{eV} (\mathcal{D})$, which is very close to the bound $\sum m_\nu \geq 0.06~\mathrm{eV}$ from the neutrino oscillation. The inclusion of RSD results in a weaker neutrino mass bound in other extended models as shown in Figs.~\ref{fig:me}, \ref{fig:w0waDE} and \ref{fig:Neff}.

Motivated by the Hubble tension, we have evaluated the impacts of RSD observations on the varying $m_e$ model, $w_0w_a$ model and $\Delta N_\mathrm{eff}$ model.
For data $\mathcal{D}+\mathrm{RSD}$, only varying $m_e$ model is compatible with a higher $H_0$, while the others fit with a smaller $H_0$ than that in $\Lambda$CDM model, because a larger EOS parameters of dark energy or smaller $N_\mathrm{eff}$ than $3$ are preferred in those models. Hence, the Hubble tension is not improved.

The best-fit points for data $\mathcal{D}+\mathrm{RSD}+\mathrm{R21}$ are
 $H_0\simeq 70~\mathrm{km/s/Mpc}$ for the $\Delta N_{\mathrm{eff}}$ model and $H_0\simeq 71~\mathrm{km/s/Mpc}$ for the varying $m_e$ model.
As far as $\Delta\chi^2_{\mathrm{RSD}}$ is concerned, 
$\chi^2_{\mathrm{RSD}}$ is worsened by about $1$ in the $\Delta N_{\mathrm{eff}}$ model and by approximately $6$ in the varying $m_e$ model, respectively.
Since the variation of $\chi^2_{R21}$ is more than those of RSD and dominate,
 the total are opposite to the response for $\chi^2_{\mathrm{RSD}}$.
This is nothing but none of those relax the Hubble tension and the amplitude of fluctuation simultaneouly.
This can be understood as follows.
To alleviate the Hubble tension by shortening the sound horizon, the value of $\sigma_8$ increases while the matter density $\Omega_m$ decreases.
In Fig.~\ref{fig:fsigma}, we plot the matter growth rate $f\sigma_8$ for the best-fit of the $\Lambda$CDM model (orange curve) and the best-fit of the varying electron mass model (blue curve).
As shown in the figure, in the low-redshift region ($0 < z < 1$), the blue curve deviates more significantly from the observational error bars compared to the orange curve.
It is the reason why the value of $\chi^2_{\mathrm{RSD}}$ gets worse in the $N_{\mathrm{eff}}$ model and the varying $m_e$ model.

\begin{figure}[ht]
\includegraphics[width=16cm]{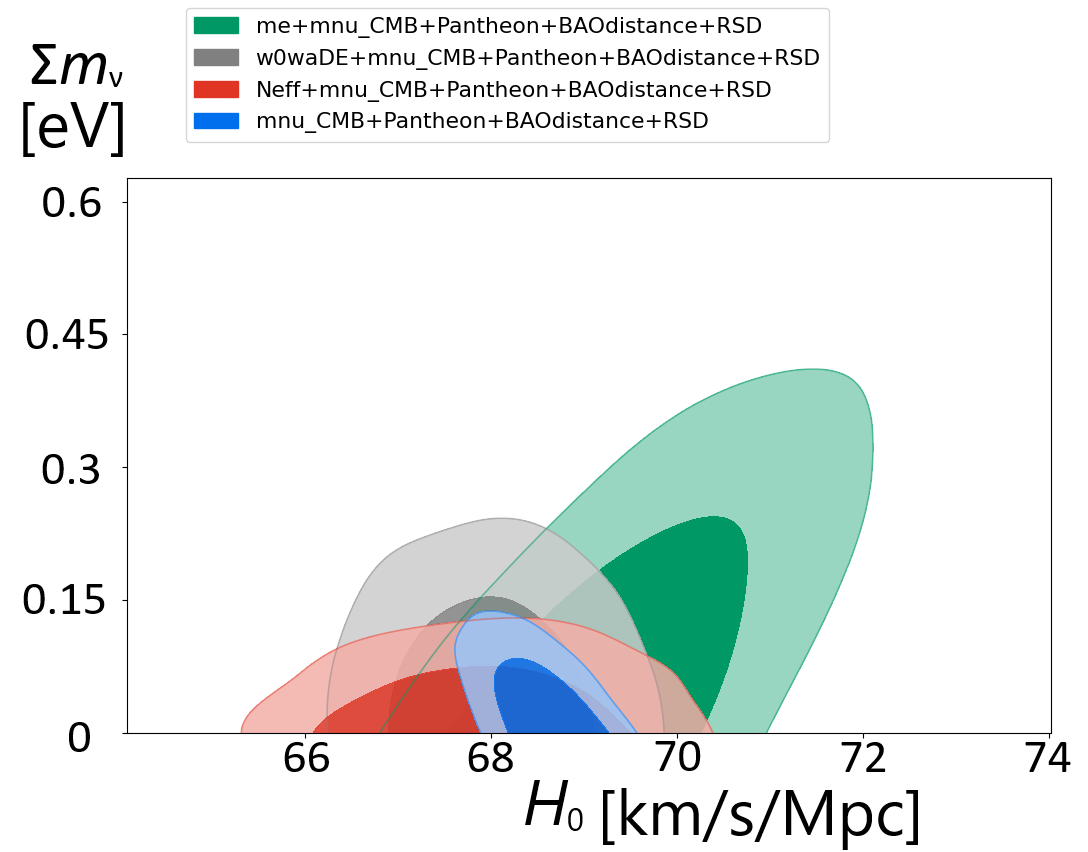}
\caption{The summary plot of the 2D constraints on $\sum m_\nu$ and $H_0$ for the data $\mathcal{D}$ with RSD.}
\label{fig:sum} 
\end{figure}

In summary, Fig.~\ref{fig:sum} shows the model comparison regarding constraints on the neutrino mass and Hubble constant, newly obtained in this paper, in the analysis with the $\mathcal{D}$+RSD data.
At a glance, it is apparent that the varying $m_e+m_\nu$ model permits both the largest neutrino mass and highest Hubble constant among the four models.

\begin{acknowledgments}
\noindent 
We are grateful to Adria Gomez-Valent for sharing the RSD Data.
This work was supported in part by JST SPRING, Grant No. JPMJSP2119 (Y.T.)
and KAKENHI Grants No.~JP23K03402 (O.S.).
 
\end{acknowledgments}

\bibliography{RSD_H0}

@article{Sekiguchi:2020teg,
    author = "Sekiguchi, Toyokazu and Takahashi, Tomo",
    title = "{Early recombination as a solution to the $H_0$ tension}",
    eprint = "2007.03381",
    archivePrefix = "arXiv",
    primaryClass = "astro-ph.CO",
    reportNumber = "KEK-TH-2238",
    doi = "10.1103/PhysRevD.103.083507",
    journal = "Phys. Rev. D",
    volume = "103",
    number = "8",
    pages = "083507",
    year = "2021"
}

@article{Kaiser:1987qv,
    author = "Kaiser, N.",
    title = "{Clustering in real space and in redshift space}",
    doi = "10.1093/mnras/227.1.1",
    journal = "Mon. Not. Roy. Astron. Soc.",
    volume = "227",
    pages = "1--27",
    year = "1987"
}

@article{Hamilton:1992zz,
    author = "Hamilton, A. J. S.",
    title = "{Measuring Omega and the real correlation function from the redshift correlation function}",
    doi = "10.1086/186264",
    journal = "Astrophys. J. Lett.",
    volume = "385",
    pages = "L5--L8",
    year = "1992"
}

@article{Rezazadeh:2022lsf,
    author = "Rezazadeh, Kazem and Ashoorioon, Amjad and Grin, Daniel",
    title = "{Cascading Dark Energy}",
    eprint = "2208.07631",
    archivePrefix = "arXiv",
    primaryClass = "astro-ph.CO",
    reportNumber = "IPM/P-2022/60",
    doi = "10.3847/1538-4357/ad7b16",
    journal = "Astrophys. J.",
    volume = "975",
    number = "1",
    pages = "137",
    year = "2024"
}

@article{Craig:2024tky,
    author = "Craig, Nathaniel and Green, Daniel and Meyers, Joel and Rajendran, Surjeet",
    title = "{No \ensuremath{\nu}s is Good News}",
    eprint = "2405.00836",
    archivePrefix = "arXiv",
    primaryClass = "astro-ph.CO",
    reportNumber = "FERMILAB-PUB-24-0492-SQMS-V",
    doi = "10.1007/JHEP09(2024)097",
    journal = "JHEP",
    volume = "09",
    pages = "097",
    year = "2024"
}

@article{Allali:2024aiv,
    author = "Allali, Itamar J. and Notari, Alessio",
    title = "{Neutrino mass bounds from DESI 2024 are relaxed by Planck PR4 and cosmological supernovae}",
    eprint = "2406.14554",
    archivePrefix = "arXiv",
    primaryClass = "astro-ph.CO",
    doi = "10.1088/1475-7516/2024/12/020",
    journal = "JCAP",
    volume = "12",
    pages = "020",
    year = "2024"
}

@article{Loverde:2024nfi,
    author = "Loverde, Marilena and Weiner, Zachary J.",
    title = "{Massive neutrinos and cosmic composition}",
    eprint = "2410.00090",
    archivePrefix = "arXiv",
    primaryClass = "astro-ph.CO",
    doi = "10.1088/1475-7516/2024/12/048",
    journal = "JCAP",
    volume = "12",
    pages = "048",
    year = "2024"
}

@article{Baryakhtar:2024rky,
    author = "Baryakhtar, Masha and Simon, Olivier and Weiner, Zachary J.",
    title = "{Cosmology with varying fundamental constants from hyperlight, coupled scalars}",
    eprint = "2405.10358",
    archivePrefix = "arXiv",
    primaryClass = "astro-ph.CO",
    doi = "10.1103/PhysRevD.110.083505",
    journal = "Phys. Rev. D",
    volume = "110",
    number = "8",
    pages = "083505",
    year = "2024"
}

@article{Song:2008qt,
    author = "Song, Yong-Seon and Percival, Will J.",
    title = "{Reconstructing the history of structure formation using Redshift Distortions}",
    eprint = "0807.0810",
    archivePrefix = "arXiv",
    primaryClass = "astro-ph",
    doi = "10.1088/1475-7516/2009/10/004",
    journal = "JCAP",
    volume = "10",
    pages = "004",
    year = "2009"
}

@article{Blake:2011rj,
    author = "Blake, Chris and others",
    title = "{The WiggleZ Dark Energy Survey: the growth rate of cosmic structure since redshift z=0.9}",
    eprint = "1104.2948",
    archivePrefix = "arXiv",
    primaryClass = "astro-ph.CO",
    doi = "10.1111/j.1365-2966.2011.18903.x",
    journal = "Mon. Not. Roy. Astron. Soc.",
    volume = "415",
    pages = "2876",
    year = "2011"
}

@article{Blake:2013nif,
    author = "Blake, Chris and others",
    title = "{Galaxy And Mass Assembly (GAMA): improved cosmic growth measurements using multiple tracers of large-scale structure}",
    eprint = "1309.5556",
    archivePrefix = "arXiv",
    primaryClass = "astro-ph.CO",
    doi = "10.1093/mnras/stt1791",
    journal = "Mon. Not. Roy. Astron. Soc.",
    volume = "436",
    pages = "3089",
    year = "2013"
}

@article{Simpson:2015yfa,
    author = "Simpson, Fergus and Blake, Chris and Peacock, John A. and Baldry, Ivan and Bland-Hawthorn, Joss and Heavens, Alan and Heymans, Catherine and Loveday, Jon and Norberg, Peder",
    title = "{Galaxy and mass assembly: Redshift space distortions from the clipped galaxy field}",
    eprint = "1505.03865",
    archivePrefix = "arXiv",
    primaryClass = "astro-ph.CO",
    doi = "10.1103/PhysRevD.93.023525",
    journal = "Phys. Rev. D",
    volume = "93",
    number = "2",
    pages = "023525",
    year = "2016"
}

@article{Okumura:2015lvp,
    author = "Okumura, Teppei and others",
    title = "{The Subaru FMOS galaxy redshift survey (FastSound). IV. New constraint on gravity theory from redshift space distortions at $z\sim 1.4$}",
    eprint = "1511.08083",
    archivePrefix = "arXiv",
    primaryClass = "astro-ph.CO",
    doi = "10.1093/pasj/psw029",
    journal = "Publ. Astron. Soc. Jap.",
    volume = "68",
    number = "3",
    pages = "38",
    year = "2016"
}

@article{Gil-Marin:2016wya,
    author = "Gil-Mar\'\i{}n, H\'ector and Percival, Will J. and Verde, Licia and Brownstein, Joel R. and Chuang, Chia-Hsun and Kitaura, Francisco-Shu and Rodr\'\i{}guez-Torres, Sergio A. and Olmstead, Matthew D.",
    title = "{The clustering of galaxies in the SDSS-III Baryon Oscillation Spectroscopic Survey: RSD measurement from the power spectrum and bispectrum of the DR12 BOSS galaxies}",
    eprint = "1606.00439",
    archivePrefix = "arXiv",
    primaryClass = "astro-ph.CO",
    doi = "10.1093/mnras/stw2679",
    journal = "Mon. Not. Roy. Astron. Soc.",
    volume = "465",
    number = "2",
    pages = "1757--1788",
    year = "2017"
}

@article{Mohammad:2018mdy,
    author = "Mohammad, F. G. and others",
    title = "{The VIMOS Public Extragalactic Redshift Survey (VIPERS): Unbiased clustering estimate with VIPERS slit assignment}",
    eprint = "1807.05999",
    archivePrefix = "arXiv",
    primaryClass = "astro-ph.CO",
    doi = "10.1051/0004-6361/201833853",
    journal = "Astron. Astrophys.",
    volume = "619",
    pages = "A17",
    year = "2018"
}

@article{Said:2020epb,
    author = "Said, Khaled and Colless, Matthew and Magoulas, Christina and Lucey, John R. and Hudson, Michael J.",
    title = "{Joint analysis of 6dFGS and SDSS peculiar velocities for the growth rate of cosmic structure and tests of gravity}",
    eprint = "2007.04993",
    archivePrefix = "arXiv",
    primaryClass = "astro-ph.CO",
    doi = "10.1093/mnras/staa2032",
    journal = "Mon. Not. Roy. Astron. Soc.",
    volume = "497",
    number = "1",
    pages = "1275--1293",
    year = "2020"
}

@article{eBOSS:2020gbb,
    author = "Hou, Jiamin and others",
    collaboration = "eBOSS",
    title = "{The Completed SDSS-IV extended Baryon Oscillation Spectroscopic Survey: BAO and RSD measurements from anisotropic clustering analysis of the Quasar Sample in configuration space between redshift 0.8 and 2.2}",
    eprint = "2007.08998",
    archivePrefix = "arXiv",
    primaryClass = "astro-ph.CO",
    doi = "10.1093/mnras/staa3234",
    journal = "Mon. Not. Roy. Astron. Soc.",
    volume = "500",
    number = "1",
    pages = "1201--1221",
    year = "2020"
}

@article{Avila:2021dqv,
    author = "Avila, F. and Bernui, A. and de Carvalho, E. and Novaes, C. P.",
    title = "{The growth rate of cosmic structures in the local Universe with the ALFALFA survey}",
    eprint = "2105.10583",
    archivePrefix = "arXiv",
    primaryClass = "astro-ph.CO",
    doi = "10.1093/mnras/stab1488",
    journal = "Mon. Not. Roy. Astron. Soc.",
    volume = "505",
    number = "3",
    pages = "3404--3413",
    year = "2021"
}

@article{Planck:2018vyg,
    author = "Aghanim, N. and others",
    collaboration = "Planck",
    title = "{Planck 2018 results. VI. Cosmological parameters}",
    eprint = "1807.06209",
    archivePrefix = "arXiv",
    primaryClass = "astro-ph.CO",
    doi = "10.1051/0004-6361/201833910",
    journal = "Astron. Astrophys.",
    volume = "641",
    pages = "A6",
    year = "2020",
    note = "[Erratum: Astron.Astrophys. 652, C4 (2021)]"
}

@article{Joudaki:2017zdt,
    author = "Joudaki, Shahab and others",
    title = "{KiDS-450 + 2dFLenS: Cosmological parameter constraints from weak gravitational lensing tomography and overlapping redshift-space galaxy clustering}",
    eprint = "1707.06627",
    archivePrefix = "arXiv",
    primaryClass = "astro-ph.CO",
    doi = "10.1093/mnras/stx2820",
    journal = "Mon. Not. Roy. Astron. Soc.",
    volume = "474",
    number = "4",
    pages = "4894--4924",
    year = "2018"
}

@article{Gomez-Valent:2017idt,
    author = "Gomez-Valent, Adria and Sola, Joan",
    title = "{Relaxing the $\sigma_8$-tension through running vacuum in the Universe}",
    eprint = "1711.00692",
    archivePrefix = "arXiv",
    primaryClass = "astro-ph.CO",
    doi = "10.1209/0295-5075/120/39001",
    journal = "EPL",
    volume = "120",
    number = "3",
    pages = "39001",
    year = "2017"
}

@article{Nesseris:2017vor,
    author = "Nesseris, Savvas and Pantazis, George and Perivolaropoulos, Leandros",
    title = "{Tension and constraints on modified gravity parametrizations of $G_{\textrm{eff}}(z)$ from growth rate and Planck data}",
    eprint = "1703.10538",
    archivePrefix = "arXiv",
    primaryClass = "astro-ph.CO",
    reportNumber = "IFT-UAM-CSIC-17-031",
    doi = "10.1103/PhysRevD.96.023542",
    journal = "Phys. Rev. D",
    volume = "96",
    number = "2",
    pages = "023542",
    year = "2017"
}

@article{Gomez-Valent:2018nib,
    author = "G\'omez-Valent, Adri\`a and Sol\`a Peracaula, Joan",
    title = "{Density perturbations for running vacuum: a successful approach to structure formation and to the $\sigma_8$-tension}",
    eprint = "1801.08501",
    archivePrefix = "arXiv",
    primaryClass = "astro-ph.CO",
    doi = "10.1093/mnras/sty1028",
    journal = "Mon. Not. Roy. Astron. Soc.",
    volume = "478",
    number = "1",
    pages = "126--145",
    year = "2018"
}

@article{Benisty:2020kdt,
    author = "Benisty, David",
    title = "{Quantifying the $S_8$ tension with the Redshift Space Distortion data set}",
    eprint = "2005.03751",
    archivePrefix = "arXiv",
    primaryClass = "astro-ph.CO",
    doi = "10.1016/j.dark.2020.100766",
    journal = "Phys. Dark Univ.",
    volume = "31",
    pages = "100766",
    year = "2021"
}

@article{Wright:2020ppw,
    author = "Wright, Angus H. and Hildebrandt, Hendrik and van den Busch, Jan Luca and Heymans, Catherine and Joachimi, Benjamin and Kannawadi, Arun and Kuijken, Konrad",
    title = "{KiDS+VIKING-450: Improved cosmological parameter constraints from redshift calibration with self-organising maps}",
    eprint = "2005.04207",
    archivePrefix = "arXiv",
    primaryClass = "astro-ph.CO",
    doi = "10.1051/0004-6361/202038389",
    journal = "Astron. Astrophys.",
    volume = "640",
    pages = "L14",
    year = "2020"
}

@article{Nunes:2021ipq,
    author = "Nunes, Rafael C. and Vagnozzi, Sunny",
    title = "{Arbitrating the S8 discrepancy with growth rate measurements from redshift-space distortions}",
    eprint = "2106.01208",
    archivePrefix = "arXiv",
    primaryClass = "astro-ph.CO",
    doi = "10.1093/mnras/stab1613",
    journal = "Mon. Not. Roy. Astron. Soc.",
    volume = "505",
    number = "4",
    pages = "5427--5437",
    year = "2021"
}

@article{Briffa:2023ozo,
    author = "Briffa, Rebecca and Escamilla-Rivera, Celia and Levi Said, Jackson and Mifsud, Jurgen",
    title = "{Growth of structures using redshift space distortion in f(T) cosmology}",
    eprint = "2310.09159",
    archivePrefix = "arXiv",
    primaryClass = "astro-ph.CO",
    doi = "10.1093/mnras/stae103",
    journal = "Mon. Not. Roy. Astron. Soc.",
    volume = "528",
    number = "2",
    pages = "2711--2727",
    year = "2024"
}

@article{Nguyen:2023fip,
    author = "Nguyen, Nhat-Minh and Huterer, Dragan and Wen, Yuewei",
    title = "{Evidence for Suppression of Structure Growth in the Concordance Cosmological Model}",
    eprint = "2302.01331",
    archivePrefix = "arXiv",
    primaryClass = "astro-ph.CO",
    reportNumber = "LCTP-23-03",
    doi = "10.1103/PhysRevLett.131.111001",
    journal = "Phys. Rev. Lett.",
    volume = "131",
    number = "11",
    pages = "111001",
    year = "2023"
}

@article{Adil:2023jtu,
    author = {Adil, Shahnawaz A. and Akarsu, \"Ozg\"ur and Malekjani, Mohammad and Colg\'ain, Eoin \'O. and Pourojaghi, Saeed and Sen, Anjan A. and Sheikh-Jabbari, M. M.},
    title = "{S8 increases with effective redshift in \ensuremath{\Lambda}CDM cosmology}",
    eprint = "2303.06928",
    archivePrefix = "arXiv",
    primaryClass = "astro-ph.CO",
    doi = "10.1093/mnrasl/slad165",
    journal = "Mon. Not. Roy. Astron. Soc.",
    volume = "528",
    number = "1",
    pages = "L20--L26",
    year = "2023"
}

@article{Tang:2024gtq,
    author = "Tang, Xin and Ma, Yin-Zhe and Dai, Wei-Ming and He, Hong-Jian",
    title = "{Constraining holographic dark energy and analyzing cosmological tensions}",
    eprint = "2407.08427",
    archivePrefix = "arXiv",
    primaryClass = "astro-ph.CO",
    doi = "10.1016/j.dark.2024.101568",
    journal = "Phys. Dark Univ.",
    volume = "46",
    pages = "101568",
    year = "2024"
}

@article{Sabogal:2024yha,
    author = "Sabogal, Miguel A. and Silva, Emanuelly and Nunes, Rafael C. and Kumar, Suresh and Di Valentino, Eleonora and Giar\`e, William",
    title = "{Quantifying the S8 tension and evidence for interacting dark energy from redshift-space distortion measurements}",
    eprint = "2408.12403",
    archivePrefix = "arXiv",
    primaryClass = "astro-ph.CO",
    doi = "10.1103/PhysRevD.110.123508",
    journal = "Phys. Rev. D",
    volume = "110",
    number = "12",
    pages = "123508",
    year = "2024"
}

@article{Toda:2024fgv,
    author = "Toda, Yo and G\'omez-Valent, Adri\`a and Koyama, Kazuya",
    title = "{Efficient Compression of Redshift-Space Distortion Data for Late-Time Modified Gravity Models}",
    eprint = "2408.16388",
    archivePrefix = "arXiv",
    primaryClass = "astro-ph.CO",
    reportNumber = "EPHOU-24-011",
    month = "8",
    year = "2024"
}

@article{Riess:2021jrx,
    author = "Riess, Adam G. and others",
    title = "{A Comprehensive Measurement of the Local Value of the Hubble Constant with 1 km s$^{−1}$ Mpc$^{−1}$ Uncertainty from the Hubble Space Telescope and the SH0ES Team}",
    eprint = "2112.04510",
    archivePrefix = "arXiv",
    primaryClass = "astro-ph.CO",
    doi = "10.3847/2041-8213/ac5c5b",
    journal = "Astrophys. J. Lett.",
    volume = "934",
    number = "1",
    pages = "L7",
    year = "2022"
}

@article{Riess:2023bfx,
    author = "Riess, Adam G. and Anand, Gagandeep S. and Yuan, Wenlong and Casertano, Stefano and Dolphin, Andrew and Macri, Lucas M. and Breuval, Louise and Scolnic, Dan and Perrin, Marshall and Anderson, Richard I.",
    title = "{Crowded No More: The Accuracy of the Hubble Constant Tested with High-resolution Observations of Cepheids by JWST}",
    eprint = "2307.15806",
    archivePrefix = "arXiv",
    primaryClass = "astro-ph.CO",
    doi = "10.3847/2041-8213/acf769",
    journal = "Astrophys. J. Lett.",
    volume = "956",
    number = "1",
    pages = "L18",
    year = "2023"
}

@article{Riess:2024ohe,
    author = "Riess, Adam G. and Anand, Gagandeep S. and Yuan, Wenlong and Casertano, Stefano and Dolphin, Andrew and Macri, Lucas M. and Breuval, Louise and Scolnic, Dan and Perrin, Marshall and Anderson, Richard I.",
    title = "{JWST Observations Reject Unrecognized Crowding of Cepheid Photometry as an Explanation for the Hubble Tension at 8\ensuremath{\sigma} Confidence}",
    eprint = "2401.04773",
    archivePrefix = "arXiv",
    primaryClass = "astro-ph.CO",
    doi = "10.3847/2041-8213/ad1ddd",
    journal = "Astrophys. J. Lett.",
    volume = "962",
    number = "1",
    pages = "L17",
    year = "2024"
}

@article{Wong:2019kwg,
    author = "Wong, Kenneth C. and others",
    title = "{H0LiCOW \textendash{} XIII. A 2.4 per cent measurement of H0 from lensed quasars: 5.3\ensuremath{\sigma} tension between early- and late-Universe probes}",
    eprint = "1907.04869",
    archivePrefix = "arXiv",
    primaryClass = "astro-ph.CO",
    doi = "10.1093/mnras/stz3094",
    journal = "Mon. Not. Roy. Astron. Soc.",
    volume = "498",
    number = "1",
    pages = "1420--1439",
    year = "2020"
}

@article{Freedman:2019jwv,
    author = "Freedman, Wendy L. and others",
    title = "{The Carnegie-Chicago Hubble Program. VIII. An Independent Determination of the Hubble Constant Based on the Tip of the Red Giant Branch}",
    eprint = "1907.05922",
    archivePrefix = "arXiv",
    primaryClass = "astro-ph.CO",
    doi = "10.3847/1538-4357/ab2f73",
    journal = "Astrophys. J.",
    volume = "882",
    number = "1",
    pages = "34",
    year = "2019"
}

@article{Freedman:2020dne,
    author = "Freedman, Wendy L. and Madore, Barry F. and Hoyt, Taylor and Jang, In Sung and Beaton, Rachael and Lee, Myung Gyoon and Monson, Andrew and Neeley, Jill and Rich, Jeffrey",
    title = "{Calibration of the Tip of the Red Giant Branch (TRGB)}",
    eprint = "2002.01550",
    archivePrefix = "arXiv",
    primaryClass = "astro-ph.GA",
    doi = "10.3847/1538-4357/ab7339",
    journal = "Astrophys. J.",
    year = {2020},
    volume = {891},
    number = {1},
    pages = {57},
}

@article{Freedman:2021ahq,
    author = "Freedman, Wendy L.",
    title = "{Measurements of the Hubble Constant: Tensions in Perspective}",
    eprint = "2106.15656",
    archivePrefix = "arXiv",
    primaryClass = "astro-ph.CO",
    doi = "10.3847/1538-4357/ac0e95",
    volume = "919",
    number = "1",
    pages = "16",
    year = "2021"
}

@article{Beutler:2011hx,
    author = "Beutler, Florian and Blake, Chris and Colless, Matthew and Jones, D.Heath and Staveley-Smith, Lister and Campbell, Lachlan and Parker, Quentin and Saunders, Will and Watson, Fred",
    title = "{The 6dF Galaxy Survey: Baryon Acoustic Oscillations and the Local Hubble Constant}",
    eprint = "1106.3366",
    archivePrefix = "arXiv",
    primaryClass = "astro-ph.CO",
    doi = "10.1111/j.1365-2966.2011.19250.x",
    journal = "Mon. Not. Roy. Astron. Soc.",
    volume = "416",
    pages = "3017--3032",
    year = "2011"
}

@article{Ross:2014qpa,
    author = "Ross, Ashley J. and Samushia, Lado and Howlett, Cullan and Percival, Will J. and Burden, Angela and Manera, Marc",
    title = "{The clustering of the SDSS DR7 main Galaxy sample -- I. A 4 per cent distance measure at $z = 0.15$}",
    eprint = "1409.3242",
    archivePrefix = "arXiv",
    primaryClass = "astro-ph.CO",
    doi = "10.1093/mnras/stv154",
    journal = "Mon. Not. Roy. Astron. Soc.",
    volume = "449",
    number = "1",
    pages = "835--847",
    year = "2015"
}

@article{BOSS:2016wmc,
    author = "Alam, Shadab and others",
    collaboration = "BOSS",
    title = "{The clustering of galaxies in the completed SDSS-III Baryon Oscillation Spectroscopic Survey: cosmological analysis of the DR12 galaxy sample}",
    eprint = "1607.03155",
    archivePrefix = "arXiv",
    primaryClass = "astro-ph.CO",
    doi = "10.1093/mnras/stx721",
    journal = "Mon. Not. Roy. Astron. Soc.",
    volume = "470",
    number = "3",
    pages = "2617--2652",
    year = "2017"
}

@article{Pan-STARRS1:2017jku,
    author = "Scolnic, D. M. and others",
    collaboration = "Pan-STARRS1",
    title = "{The Complete Light-curve Sample of Spectroscopically Confirmed SNe Ia from Pan-STARRS1 and Cosmological Constraints from the Combined Pantheon Sample}",
    eprint = "1710.00845",
    archivePrefix = "arXiv",
    primaryClass = "astro-ph.CO",
    doi = "10.3847/1538-4357/aab9bb",
    journal = "Astrophys. J.",
    volume = "859",
    number = "2",
    pages = "101",
    year = "2018"
}

@article{Brout:2022vxf,
    author = "Brout, Dillon and others",
    title = "{The Pantheon+ Analysis: Cosmological Constraints}",
    eprint = "2202.04077",
    archivePrefix = "arXiv",
    primaryClass = "astro-ph.CO",
    doi = "10.3847/1538-4357/ac8e04",
    journal = "Astrophys. J.",
    volume = "938",
    number = "2",
    pages = "110",
    year = "2022"
}

@article{eBOSS:2020yzd,
    author = "Alam, Shadab and others",
    collaboration = "eBOSS",
    title = "{Completed SDSS-IV extended Baryon Oscillation Spectroscopic Survey: Cosmological implications from two decades of spectroscopic surveys at the Apache Point Observatory}",
    eprint = "2007.08991",
    archivePrefix = "arXiv",
    primaryClass = "astro-ph.CO",
    doi = "10.1103/PhysRevD.103.083533",
    journal = "Phys. Rev. D",
    volume = "103",
    number = "8",
    pages = "083533",
    year = "2021"
}

@article{DiValentino:2021izs,
    author = "Di Valentino, Eleonora and Mena, Olga and Pan, Supriya and Visinelli, Luca and Yang, Weiqiang and Melchiorri, Alessandro and Mota, David F. and Riess, Adam G. and Silk, Joseph",
    title = "{In the realm of the Hubble tension\textemdash{}a review of solutions}",
    eprint = "2103.01183",
    archivePrefix = "arXiv",
    primaryClass = "astro-ph.CO",
    reportNumber = "IPPP/20/108",
    doi = "10.1088/1361-6382/ac086d",
    journal = "Class. Quant. Grav.",
    volume = "38",
    number = "15",
    pages = "153001",
    year = "2021"
}

@article{Perivolaropoulos:2021jda,
    author = "Perivolaropoulos, Leandros and Skara, Foteini",
    title = "{Challenges for \ensuremath{\Lambda}CDM: An update}",
    eprint = "2105.05208",
    archivePrefix = "arXiv",
    primaryClass = "astro-ph.CO",
    doi = "10.1016/j.newar.2022.101659",
    journal = "New Astron. Rev.",
    volume = "95",
    pages = "101659",
    year = "2022"
}

@article{Schoneberg:2021qvd,
    author = {Sch\"oneberg, Nils and Franco Abell\'an, Guillermo and P\'erez S\'anchez, Andrea and Witte, Samuel J. and Poulin, Vivian and Lesgourgues, Julien},
    title = "{The H0 Olympics: A fair ranking of proposed models}",
    eprint = "2107.10291",
    archivePrefix = "arXiv",
    primaryClass = "astro-ph.CO",
    doi = "10.1016/j.physrep.2022.07.001",
    journal = "Phys. Rept.",
    volume = "984",
    pages = "1--55",
    year = "2022"
}

@article{Shah:2021onj,
    author = "Shah, Paul and Lemos, Pablo and Lahav, Ofer",
    title = "{A buyer\textquoteright{}s guide to the Hubble constant}",
    eprint = "2109.01161",
    archivePrefix = "arXiv",
    primaryClass = "astro-ph.CO",
    doi = "10.1007/s00159-021-00137-4",
    journal = "Astron. Astrophys. Rev.",
    volume = "29",
    number = "1",
    pages = "9",
    year = "2021"
}

@article{Abdalla:2022yfr,
    author = "Abdalla, Elcio and others",
    title = "{Cosmology intertwined: A review of the particle physics, astrophysics, and cosmology associated with the cosmological tensions and anomalies}",
    eprint = "2203.06142",
    archivePrefix = "arXiv",
    primaryClass = "astro-ph.CO",
    reportNumber = "FERMILAB-CONF-22-192-SCD",
    doi = "10.1016/j.jheap.2022.04.002",
    journal = "JHEAp",
    volume = "34",
    pages = "49--211",
    year = "2022"
}

@article{Hu:2023jqc,
    author = "Hu, Jian-Ping and Wang, Fa-Yin",
    title = "{Hubble Tension: The Evidence of New Physics}",
    eprint = "2302.05709",
    archivePrefix = "arXiv",
    primaryClass = "astro-ph.CO",
    doi = "10.3390/universe9020094",
    journal = "Universe",
    volume = "9",
    number = "2",
    pages = "94",
    year = "2023"
}

@article{Poulin:2018cxd,
    author = "Poulin, Vivian and Smith, Tristan L. and Karwal, Tanvi and Kamionkowski, Marc",
    title = "{Early Dark Energy Can Resolve The Hubble Tension}",
    eprint = "1811.04083",
    archivePrefix = "arXiv",
    primaryClass = "astro-ph.CO",
    doi = "10.1103/PhysRevLett.122.221301",
    journal = "Phys. Rev. Lett.",
    volume = "122",
    number = "22",
    pages = "221301",
    year = "2019"
}

@article{Hart:2019dxi,
    author = "Hart, Luke and Chluba, Jens",
    title = "{Updated fundamental constant constraints from Planck 2018 data and possible relations to the Hubble tension}",
    eprint = "1912.03986",
    archivePrefix = "arXiv",
    primaryClass = "astro-ph.CO",
    doi = "10.1093/mnras/staa412",
    journal = "Mon. Not. Roy. Astron. Soc.",
    volume = "493",
    number = "3",
    pages = "3255--3263",
    year = "2020"
}

@article{Vagnozzi:2019ezj,
    author = "Vagnozzi, Sunny",
    title = "{New physics in light of the $H_0$ tension: An alternative view}",
    eprint = "1907.07569",
    archivePrefix = "arXiv",
    primaryClass = "astro-ph.CO",
    doi = "10.1103/PhysRevD.102.023518",
    journal = "Phys. Rev. D",
    volume = "102",
    number = "2",
    pages = "023518",
    year = "2020"
}

@article{Planck:2014ylh,
    author = "Ade, P. A. R. and others",
    collaboration = "Planck",
    title = "{Planck intermediate results - XXIV. Constraints on variations in fundamental constants}",
    eprint = "1406.7482",
    archivePrefix = "arXiv",
    primaryClass = "astro-ph.CO",
    doi = "10.1051/0004-6361/201424496",
    journal = "Astron. Astrophys.",
    volume = "580",
    pages = "A22",
    year = "2015"
}

@article{Solomon:2022qqf,
    author = "Solomon, Rance and Agarwal, Garvita and Stojkovic, Dejan",
    title = "{Environment dependent electron mass and the Hubble constant tension}",
    eprint = "2201.03127",
    archivePrefix = "arXiv",
    primaryClass = "hep-ph",
    doi = "10.1103/PhysRevD.105.103536",
    journal = "Phys. Rev. D",
    volume = "105",
    number = "10",
    pages = "103536",
    year = "2022"
}

@article{Hoshiya:2022ady,
    author = "Hoshiya, Kouki and Toda, Yo",
    title = "{Electron mass variation from dark sector interactions and compatibility with cosmological observations}",
    eprint = "2202.07714",
    archivePrefix = "arXiv",
    primaryClass = "astro-ph.CO",
    reportNumber = "EPHOU-22-005",
    doi = "10.1103/PhysRevD.107.043505",
    journal = "Phys. Rev. D",
    volume = "107",
    number = "4",
    pages = "043505",
    year = "2023"
}

@article{Barrow:2005qf,
    author = "Barrow, John D. and Magueijo, Joao",
    title = "{Cosmological constraints on a dynamical electron mass}",
    eprint = "astro-ph/0503222",
    archivePrefix = "arXiv",
    doi = "10.1103/PhysRevD.72.043521",
    journal = "Phys. Rev. D",
    volume = "72",
    pages = "043521",
    year = "2005"
}

@article{DESI:2024mwx,
    author = "Adame, A. G. and others",
    collaboration = "DESI",
    title = "{DESI 2024 VI: Cosmological Constraints from the Measurements of Baryon Acoustic Oscillations}",
    eprint = "2404.03002",
    archivePrefix = "arXiv",
    primaryClass = "astro-ph.CO",
    reportNumber = "FERMILAB-PUB-24-0154-PPD",
    month = "4",
    year = "2024"
}

@article{DES:2021wwk,
    author = "Abbott, T. M. C. and others",
    collaboration = "DES",
    title = "{Dark Energy Survey Year 3 results: Cosmological constraints from galaxy clustering and weak lensing}",
    eprint = "2105.13549",
    archivePrefix = "arXiv",
    primaryClass = "astro-ph.CO",
    reportNumber = "FERMILAB-PUB-21-221-AE, DES-2020-0617",
    doi = "10.1103/PhysRevD.105.023520",
    journal = "Phys. Rev. D",
    volume = "105",
    number = "2",
    pages = "023520",
    year = "2022"
}

@article{Lewis:2002ah,
    author = "Lewis, Antony and Bridle, Sarah",
    title = "{Cosmological parameters from CMB and other data: A Monte Carlo approach}",
    eprint = "astro-ph/0205436",
    archivePrefix = "arXiv",
    doi = "10.1103/PhysRevD.66.103511",
    journal = "Phys. Rev. D",
    volume = "66",
    pages = "103511",
    year = "2002"
}

@article{Planck:2018lbu,
    author = "Aghanim, N. and others",
    collaboration = "Planck",
    title = "{Planck 2018 results. VIII. Gravitational lensing}",
    eprint = "1807.06210",
    archivePrefix = "arXiv",
    primaryClass = "astro-ph.CO",
    doi = "10.1051/0004-6361/201833886",
    journal = "Astron. Astrophys.",
    volume = "641",
    pages = "A8",
    year = "2020"
}

@article{Seto:2022xgx,
    author = "Seto, Osamu and Toda, Yo",
    title = "{Big bang nucleosynthesis constraints on varying electron mass solution to the Hubble tension}",
    eprint = "2206.13209",
    archivePrefix = "arXiv",
    primaryClass = "astro-ph.CO",
    reportNumber = "EPHOU-22-011",
    doi = "10.1103/PhysRevD.107.083512",
    journal = "Phys. Rev. D",
    volume = "107",
    number = "8",
    pages = "083512",
    year = "2023"
}

@article{Seto:2021xua,
    author = "Seto, Osamu and Toda, Yo",
    title = "{Comparing early dark energy and extra radiation solutions to the Hubble tension with BBN}",
    eprint = "2101.03740",
    archivePrefix = "arXiv",
    primaryClass = "astro-ph.CO",
    reportNumber = "EPHOU-21-002",
    doi = "10.1103/PhysRevD.103.123501",
    journal = "Phys. Rev. D",
    volume = "103",
    number = "12",
    pages = "123501",
    year = "2021"
}

@article{Poulin:2018dzj,
    author = "Poulin, Vivian and Smith, Tristan L. and Grin, Daniel and Karwal, Tanvi and Kamionkowski, Marc",
    title = "{Cosmological implications of ultralight axionlike fields}",
    eprint = "1806.10608",
    archivePrefix = "arXiv",
    primaryClass = "astro-ph.CO",
    doi = "10.1103/PhysRevD.98.083525",
    journal = "Phys. Rev. D",
    volume = "98",
    number = "8",
    pages = "083525",
    year = "2018"
}

@article{Braglia:2020bym,
    author = "Braglia, Matteo and Emond, William T. and Finelli, Fabio and Gumrukcuoglu, A. Emir and Koyama, Kazuya",
    title = "{Unified framework for early dark energy from $\alpha$-attractors}",
    eprint = "2005.14053",
    archivePrefix = "arXiv",
    primaryClass = "astro-ph.CO",
    doi = "10.1103/PhysRevD.102.083513",
    journal = "Phys. Rev. D",
    volume = "102",
    number = "8",
    pages = "083513",
    year = "2020"
}

@article{Agrawal:2019lmo,
    author = "Agrawal, Prateek and Cyr-Racine, Francis-Yan and Pinner, David and Randall, Lisa",
    title = "{Rock \textquoteleft{}n\textquoteright{} roll solutions to the Hubble tension}",
    eprint = "1904.01016",
    archivePrefix = "arXiv",
    primaryClass = "astro-ph.CO",
    doi = "10.1016/j.dark.2023.101347",
    journal = "Phys. Dark Univ.",
    volume = "42",
    pages = "101347",
    year = "2023"
}

@article{Ye:2020btb,
    author = "Ye, Gen and Piao, Yun-Song",
    title = "{Is the Hubble tension a hint of AdS phase around recombination?}",
    eprint = "2001.02451",
    archivePrefix = "arXiv",
    primaryClass = "astro-ph.CO",
    doi = "10.1103/PhysRevD.101.083507",
    journal = "Phys. Rev. D",
    volume = "101",
    number = "8",
    pages = "083507",
    year = "2020"
}

@article{Smith:2019ihp,
    author = "Smith, Tristan L. and Poulin, Vivian and Amin, Mustafa A.",
    title = "{Oscillating scalar fields and the Hubble tension: a resolution with novel signatures}",
    eprint = "1908.06995",
    archivePrefix = "arXiv",
    primaryClass = "astro-ph.CO",
    doi = "10.1103/PhysRevD.101.063523",
    journal = "Phys. Rev. D",
    volume = "101",
    number = "6",
    pages = "063523",
    year = "2020"
}

@article{Lin:2019qug,
    author = "Lin, Meng-Xiang and Benevento, Giampaolo and Hu, Wayne and Raveri, Marco",
    title = "{Acoustic Dark Energy: Potential Conversion of the Hubble Tension}",
    eprint = "1905.12618",
    archivePrefix = "arXiv",
    primaryClass = "astro-ph.CO",
    doi = "10.1103/PhysRevD.100.063542",
    journal = "Phys. Rev. D",
    volume = "100",
    number = "6",
    pages = "063542",
    year = "2019"
}

@article{Niedermann:2019olb,
    author = "Niedermann, Florian and Sloth, Martin S.",
    title = "{New early dark energy}",
    eprint = "1910.10739",
    archivePrefix = "arXiv",
    primaryClass = "astro-ph.CO",
    doi = "10.1103/PhysRevD.103.L041303",
    journal = "Phys. Rev. D",
    volume = "103",
    number = "4",
    pages = "L041303",
    year = "2021"
}

@article{Niedermann:2020dwg,
    author = "Niedermann, Florian and Sloth, Martin S.",
    title = "{Resolving the Hubble tension with new early dark energy}",
    eprint = "2006.06686",
    archivePrefix = "arXiv",
    primaryClass = "astro-ph.CO",
    doi = "10.1103/PhysRevD.102.063527",
    journal = "Phys. Rev. D",
    volume = "102",
    number = "6",
    pages = "063527",
    year = "2020"
}

@article{Hart:2017ndk,
    author = "Hart, Luke and Chluba, Jens",
    title = "{New constraints on time-dependent variations of fundamental constants using Planck data}",
    eprint = "1705.03925",
    archivePrefix = "arXiv",
    primaryClass = "astro-ph.CO",
    doi = "10.1093/mnras/stx2783",
    journal = "Mon. Not. Roy. Astron. Soc.",
    volume = "474",
    number = "2",
    pages = "1850--1861",
    year = "2018"
}

@article{Allali:2024cji,
    author = "Allali, Itamar J. and Notari, Alessio and Rompineve, Fabrizio",
    title = "{Dark Radiation with Baryon Acoustic Oscillations from DESI 2024 and the $H_0$ tension}",
    eprint = "2404.15220",
    archivePrefix = "arXiv",
    primaryClass = "astro-ph.CO",
    month = "4",
    year = "2024"
}

@article{Chevallier:2000qy,
    author = "Chevallier, Michel and Polarski, David",
    title = "{Accelerating universes with scaling dark matter}",
    eprint = "gr-qc/0009008",
    archivePrefix = "arXiv",
    doi = "10.1142/S0218271801000822",
    journal = "Int. J. Mod. Phys. D",
    volume = "10",
    pages = "213--224",
    year = "2001"
}

@article{Linder:2002et,
    author = "Linder, Eric V.",
    title = "{Exploring the expansion history of the universe}",
    eprint = "astro-ph/0208512",
    archivePrefix = "arXiv",
    doi = "10.1103/PhysRevLett.90.091301",
    journal = "Phys. Rev. Lett.",
    volume = "90",
    pages = "091301",
    year = "2003"
}

@article{Tsujikawa:2012hv,
    author = "Tsujikawa, Shinji and De Felice, Antonio and Alcaniz, Jailson",
    title = "{Testing for dynamical dark energy models with redshift-space distortions}",
    eprint = "1210.4239",
    archivePrefix = "arXiv",
    primaryClass = "astro-ph.CO",
    doi = "10.1088/1475-7516/2013/01/030",
    journal = "JCAP",
    volume = "01",
    pages = "030",
    year = "2013"
}

@article{Alestas:2020mvb,
    author = "Alestas, G. and Kazantzidis, L. and Perivolaropoulos, L.",
    title = "{$H_0$ tension, phantom dark energy, and cosmological parameter degeneracies}",
    eprint = "2004.08363",
    archivePrefix = "arXiv",
    primaryClass = "astro-ph.CO",
    doi = "10.1103/PhysRevD.101.123516",
    journal = "Phys. Rev. D",
    volume = "101",
    number = "12",
    pages = "123516",
    year = "2020"
}

@article{Martinelli:2019krf,
    author = "Martinelli, Matteo and Tutusaus, Isaac",
    title = "{CMB tensions with low-redshift $H_0$ and $S_8$ measurements: impact of a redshift-dependent type-Ia supernovae intrinsic luminosity}",
    eprint = "1906.09189",
    archivePrefix = "arXiv",
    primaryClass = "astro-ph.CO",
    doi = "10.3390/sym11080986",
    journal = "Symmetry",
    volume = "11",
    number = "8",
    pages = "986",
    year = "2019"
}

@article{Seto:2024cgo,
    author = "Seto, Osamu and Toda, Yo",
    title = "{DESI constraints on the varying electron mass model and axionlike early dark energy}",
    eprint = "2405.11869",
    archivePrefix = "arXiv",
    primaryClass = "astro-ph.CO",
    reportNumber = "EPHOU-24-006",
    doi = "10.1103/PhysRevD.110.083501",
    journal = "Phys. Rev. D",
    volume = "110",
    number = "8",
    pages = "083501",
    year = "2024"
}

@article{Toda:2024ncp,
    author = {Toda, Yo and Giar\`e, William and \"Oz\"ulker, Emre and Di Valentino, Eleonora and Vagnozzi, Sunny},
    title = "{Combining pre- and post-recombination new physics to address cosmological tensions: Case study with varying electron mass and sign-switching cosmological constant}",
    eprint = "2407.01173",
    archivePrefix = "arXiv",
    primaryClass = "astro-ph.CO",
    doi = "10.1016/j.dark.2024.101676",
    journal = "Phys. Dark Univ.",
    volume = "46",
    pages = "101676",
    year = "2024"
}

@article{Braglia:2020auw,
    author = "Braglia, Matteo and Ballardini, Mario and Finelli, Fabio and Koyama, Kazuya",
    title = "{Early modified gravity in light of the $H_0$ tension and LSS data}",
    eprint = "2011.12934",
    archivePrefix = "arXiv",
    primaryClass = "astro-ph.CO",
    doi = "10.1103/PhysRevD.103.043528",
    journal = "Phys. Rev. D",
    volume = "103",
    number = "4",
    pages = "043528",
    year = "2021"
}

@article{Seto:2021tad,
    author = "Seto, Osamu and Toda, Yo",
    title = "{Hubble tension in lepton asymmetric cosmology with an extra radiation}",
    eprint = "2104.04381",
    archivePrefix = "arXiv",
    primaryClass = "astro-ph.CO",
    reportNumber = "EPHOU-21-008",
    doi = "10.1103/PhysRevD.104.063019",
    journal = "Phys. Rev. D",
    volume = "104",
    number = "6",
    pages = "063019",
    year = "2021"
}

\end{document}